\documentclass[twocolumn,eqsecnum,showpacs,showkeys,amsmath,amssymb,nofootinbib,superscriptaddress,floatfix]{revtex4}
\usepackage{graphicx}
\usepackage{subfigure}
\usepackage{bm}

\def\vec#1{\mathchoice{\mbox{\boldmath$\displaystyle#1$}}
{\mbox{\boldmath$\textstyle#1$}}
{\mbox{\boldmath$\scriptstyle#1$}}
{\mbox{\boldmath$\scriptscriptstyle#1$}}}
\makeatletter
\newcommand\erfc{\mathop{\operator@font erfc}\nolimits}
\def\slashchar#1{\setbox0=\hbox{$#1$}
   \dimen0=\wd0 \setbox1=\hbox{/} \dimen1=\wd1
   \ifdim\dimen0>\dimen1 \rlap{\hbox to \dimen0{\hfil/\hfil}} #1
   \else  \rlap{\hbox to \dimen1{\hfil$#1$\hfil}} / \fi}

\makeatother

\begin{document}
\title{ Bulk and shear viscosities  
of  matter created in relativistic heavy-ion collisions}
\author{Piotr Bo\.zek}
\email{Piotr.Bozek@ifj.edu.pl}
\affiliation{The H. Niewodnicza\'nski Institute of Nuclear Physics,
PL-31342 Krak\'ow, Poland} \affiliation{
Institute of Physics, Rzesz\'ow University, 
PL-35959 Rzesz\'ow, Poland}
\date{\today}

\begin{abstract}
We study the effects of  shear and bulk viscosities in the 
hadronic phase on the expansion of the fireball and on the 
particle production in relativistic heavy ion collisions. 
Comparing simulation with or without viscosity in the hadronic
 matter  we find that elliptic flow observables 
strongly dependent on dissipative effects in the late stage. 
On the other hand, interferometry radii are sensitive, 
through the early transverse flow, on the value of the 
viscosity at high temperatures. We present first calculations
 including the effects of bulk viscosity in the hadronic phase and in the hadron
emission. We find them important in obtaining a  small freeze-out temperature
consistent with the measured transverse momentum spectra and 
 elliptic flow of identified particles.
\end{abstract}

\pacs{25.75.-q, 25.75.Dw, 25.75.Ld}

\keywords{relativistic 
heavy-ion collisions, viscous  hydrodynamic model, collective flow}

\maketitle

\section{Introduction}

The matter created in relativistic heavy-ion collisions at
 the BNL Relativistic Heavy Ion Collider (RHIC) is a dense strongly 
interacting fluid \cite{Arsene:2004fa,Back:2004je,Adams:2005dq,Adcox:2004mh}. 
The observation of  strong collective
 transverse and elliptic flows is an indication that the system behaves
as a fluid. To model the dynamics of such a system
 relativistic hydrodynamics of a perfect fluid has been successfully 
applied
\cite{Teaney:2000cw,Kolb:2003dz,Hirano:2002ds,Hama:2005dz,Huovinen:2006jp,Hirano:2005xf,Broniowski:2008vp}.
The fireball expands and cools down, until some freeze-out temperature 
is reached, after which particles are emitted from a freeze-out hypersurface.
Final particle spectra, to be compared with experimental data, are obtained 
after resonance decays.
Transverse momentum spectra in the
 azimuthal angle at central rapidity are written
as an expansion in Fourier coefficients
\begin{equation}
\frac{dN}{d^2p_\perp dy}=\frac{dN}
{2\pi p_\perp dp_\perp dy}\left(1+v_2 \cos(2\phi)+\dots \right) \ .
\end{equation}
The form of the observed transverse momentum spectra 
$\frac{dN}{2\pi p_\perp dp_\perp dy}$ and the elliptic flow 
coefficient $v_2$ can be described using a convolution of the thermal 
 emission of particles with the collective velocity of the fluid itself 
\cite{Schnedermann:1993ws,Kolb:2000sd}.

Due to the rapid expansion of the dense system created in relativistic 
heavy-ion collisions deviations from local equilibrium can be important. For
 the modeling of the 
expansion of the fireball it means that viscous relativistic hydrodynamics 
should be used \cite{Muronga:2001zk,Teaney:2003kp,Baier:2006gy,Romatschke:2007mq,Chaudhuri:2006jd,Song:2007fn}.
 A consistent causal scheme 
requires the use of second order viscous equations \cite{IS}. 
Most of the existing 
applications of viscous hydrodynamics in heavy-ion collisions
 consider shear viscosity only. The value of the ratio $\eta/s$
 of the shear viscosity
 coefficient to the entropy density is an important characteristic of the 
strongly interacting medium created in the course of the collision 
\cite{Kovtun:2004de,Csernai:2006zz}. 
The extraction of the shear viscosity coefficient from the 
measured elliptic flow could give valuable information
 \cite{Romatschke:2007mq,Song:2008hj}. 
%The elliptic flow of charged
% particles as function transverse momentum from viscous hydrodynamic 
%calculations at different $\eta/s$ values  is compared to the 
%data. The estimated value of $\eta/s=1$-$2\times\frac{1}{4\pi}$ is close
% to the conjectured 
%lower bound for the most perfect fluids \cite{Kovtun:2004de}.
 The main source
 of uncertainty in the analysis lies in  the assumption of the 
initial eccentricity of the source at a given impact parameter 
\cite{Song:2008hj}.
 The causes influencing   the initial 
shape of the source include~:
different underlying models of the initial density, Color Glass
 Condensate or Glauber Model  \cite{Drescher:2006ca},
the inclusion of  binary collision contributions \cite{Kolb:2001qz},
  possible 
 initial fluctuations  of the shape (standard versus participant eccentricity)
\cite{Miller:2003kd,Alver:2007rm} 
or  a core-corona effect, where only the 
dense part of the source evolves collectively  \cite{Bozek:2008zw}.
The elliptic flow of the bulk of the matter is generated in the early 
stages of the collision. However, 
the final elliptic flow of observed hadrons is modified in the 
hadron gas 
 phase of the expansion, due to rescattering and resonance decays 
\cite{Hirano:2005wx,Hirano:2000eu,Hirano:2007xd,Hirano:2005xf}.  
This is true  both for the 
 elliptic flow 
of charged particles as well as of identified particles 
\cite{Huovinen:2005gy,Hirano:2005xf}. 
In particular,  to reproduce the observed 
splitting  between 
  pions and protons in the transverse momentum dependence of $v_2$ 
a late freeze-out or a hadronic cascade stage are
 required in the evolution.

 The role of  dissipation in 
the hadronic phase must be assessed before a reliable estimate of  viscosity 
in the (quark-gluon plasma)
QGP phase can be made.  Although the importance of the difference of the 
viscosity coefficients  in the hadronic and plasma phases has been discussed
 \cite{Hirano:2005wx}, most of the existing hydrodynamic simulations applied
 to heavy ion collisions use a constant $\eta/s$ ratio through the evolution.
In this paper we study the effect of viscosity 
in the hydrodynamic evolution below the transition temperature on the final 
elliptic flow, spectra and Hanburry Brown-Twiss (HBT) correlation radii. In 
particular  we analyze the observable differences 
in soft momenta observables between two extreme assumptions 
on the shear viscosity in the plasma 
phase ($\eta/s=0$ or $0.16$), 
after hadronic dissipation is taken into account. 
We show that the
 effect of dissipation in the hadronic phase strongly reduces the 
sensitivity of the elliptic flow measure on the value of the 
viscosity in the early
 QGP phase of the expansion. If the hadronic phase in  the expansion 
is dilute enough,
a cascade after burner can used after an early   freeze-out of the fluid 
\cite{Bass:2000ib,Nonaka:2006yn,Hirano:2007xd,Werner:2009fa}.
 Alternatively a longer hydrodynamic evolution can be used with 
a hadronic equation of state below the transition temperature.  This 
paper 
studies the effect of such a  longer hydrodynamic evolution in 
the hadronic phase using viscous hydrodynamics. % Shear and bulk viscosities
% could be very large in the hadronic phase 
%\cite{Demir:2008tr,NoronhaHostler:2008ju}, with  estimates depending 
%on the microscopic
% processes involved and the number of available hadronic states. 
 We use a moderate   value of $\eta/s=0.1$ 
in the  hadronic phase and a bulk viscosity $\zeta/s=0.03$-$0.04$. 
We show that even such small values of viscosities in the hydrodynamic 
evolution in the 
late phase of the collision   are important for the final elliptic flow, 
and that with such assumptions we  can 
 reproduce  many  experimental observations.
We calculate also the HBT radii after a hydrodynamic evolution with 
different viscosities in the QGP and hadron gas phases.

\section{Shear and bulk viscosities}

\label{sec:relax}

\begin{figure}
\includegraphics[width=.37\textwidth]{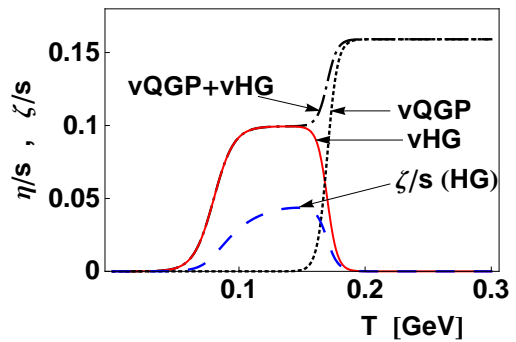}
\caption{(Color online) Temperature dependence of the ratio of shear 
and bulk viscosities to  the entropy. The solid, dotted and dash-dotted
 lines represent the shear viscosity for the vHG, vQGP and vQGP+vHG scenarios.
 The dashed line represent the bulk viscosity, effective only in the 
vHG and vQGP+vHG scenarios.}
\label{fig:etas}
\end{figure}

Besides the ideal fluid expansion we consider three other scenarios for 
the shear and bulk viscosities in the hot matter.
The general idea is that the shear viscosity in the hadronic and QGP
 phases could be very different.
Moreover if the shear viscosity in the 
hadronic phase is non-zero, it could be accompanied by non-negligible
 bulk viscosity. 
The  formula for the temperature dependence of the ratio of the 
shear viscosity to the entropy is taken in the form
\begin{equation}
\frac{\eta}{s}(T)=f_{low}(T)\frac{\eta_{HG}}{s} f_{HG}(T)+(1-f_{HG}(T))\frac{\eta_{QGP}}{s}
\end{equation}
where the function $f_{HG}(T)=1/\left(\exp\left((T-T_c)/\Delta T\right)+1\right)$
cuts-off the hadron gas viscosity above $T_c=170$MeV ( $\Delta T=8$MeV ).
$f_{low}(T)=1/\left(\exp\left((T_{low}-T)/\Delta T\right)+1\right)$ is 
introduced  to cut-off viscosity effects below $T_{low}=80$MeV  to
 improve numerical stability.
Depending on the chosen values of the viscosities
 in the hadronic matter and in QGP we consider four different scenarios
 (Table \ref{table}).
The temperature dependence of the viscosities corresponding
 to viscous scenarios in
 the Table  are shown in Fig. \ref{fig:etas}.
The scenario denoted as {vHG} assumes that only viscosity in the 
hadronic phase is non-zero. { vQGP} is taken for illustration to show 
how the dissipative phenomena in the plasma alone could influence 
the final observables. The scenario { vQGP+vHG} is the most general 
(with a suitable choice of $\eta_{HG}$ and $\eta_{QGP}$). 
The two scenarios { vHG} and { vQGH+vHG} differ by the 
choice of the shear viscosity coefficient in the plasma. The comparison 
 of these two different scenarios is one of the motivations of 
this investigation, namely to test how sensitive the final observables are
 to the assumed viscosity in the plasma phase
 (from $\eta/s=0$ to $\eta/s=0.16$), when
 dissipation in the hadronic phase occurs afterwards. These two scenarios
scenarios represent two extreme assumptions on the temperature dependence
of the ratio $\eta/s$, i.e. increasing or decreasing 
when switching from the  QGP to hadronic 
matter. Existing viscous hydrodynamic simulations assume
 a constant $\eta/s$ as function of temperature, 
microscopic estimates suggest
a (strong) increase of $\eta/s$ when decreasing the temperature. 
We test  a scenario with a moderate increase of $\eta/s$ at $T_c$  and
also another
 extreme scenario where the reverse happens 
and show that the results are in fact very similar
and that reproducing experimental data requires a small value of $\eta_{HG}$ 
for any QGP viscosity.

\begin{table}
\begin{tabular}{|c|c|c|c|}
\hline
acronym & $\frac{eta_{HG}}{s}$ & $\frac{\eta_{QGP}}{s}$& $T_F$ (MeV)\\
\hline
{ id. fl.} & 0 & 0 & 140 \\
{ vHG} & 0.1 & 0 & 150 \\
{ vQGP} & 0 & 0.16 & 130 \\
{ vHG+vQGP} & 0.1 & 0.16 & 135 \\
\hline
\end{tabular}
\caption{
 Viscosity parameters used in the four calculations presented in the paper. The last column contains the freeze-out temperature that reproduces best pion spectra in each case. }
\label{table}
\end{table}

The near equilibrium regime in a  dilute gas of interacting hadrons can be
described using the Boltzmann equation. Estimates of shear viscosity  
with hadronic cross sections or chiral models
 lead to a large value 
$\eta/s\simeq 1$  for temperatures $T\simeq 150$MeV
 \cite{Prakash:1993bt,Demir:2008tr,Dobado:2009ek,Chen:2006iga,Itakura:2007mx}.
The large value
$\eta/s \simeq 1$ in the hadronic phase seems to contradict existing fits 
of the data using viscous hydrodynamics, where $\eta/s=0.08$-$0.16$ depending 
on the initial eccentricity \cite{Romatschke:2007mq}. Also such a large value
 of the viscosity coefficient would simply mean that the 
viscous hydrodynamics cannot be applied.
Shear viscosity could be significantly reduced if the number of hadronic
 states increases near $T_c$ \cite{NoronhaHostler:2008ju}.
At temperatures close to the transition temperature the description 
of the dense medium using a transport equation approach involving hadrons with 
vacuum properties could break down. On the other hand, in microscopic models
the bulk 
viscosity is estimated to be much smaller $\zeta/s\simeq 0.03$-$0.05$ 
\cite{FernandezFraile:2008vu,NoronhaHostler:2008ju}.

In this paper, we use 
relativistic viscous hydrodynamics to model the dense hot 
medium on the low temperature side of the transition temperature.
The equation of state of  matter 
for $T<T_c$ is approximated as the hadron gas  equation of state 
involving $371$ known hadrons and resonances \cite{Chojnacki:2007jc}. This 
equation of state can be smoothly connected to the equation of state 
calculated
 in lattice QCD at higher temperatures. The final equation of state
 leads to a correct 
description of spectra and HBT radii in ideal fluid hydrodynamics
 \cite{Broniowski:2008vp,Chojnacki:2007rq}. Shear viscosity in the 
hadronic phase is 
treated as a free parameter in our calculation.
A simple estimate of the viscosities can be obtained in the 
relaxation time approximation \cite{Hosoya:1983xm,Gavin:1985ph,Sasaki:2008fg}.
Starting from the Boltzmann equation for the phase space distribution
 distribution $f_n$ of particle species $n$
\begin{equation}
p^\mu\partial_\mu f_n=-\frac{ p^\mu u_\mu(x) \delta f_n }{\tau_{HG}}
\label{eq:boltz}
\end{equation}
where $\delta f_n=f_n-f_n^0$ is the deviation from the equilibrium 
(Bose-Einstein or Fermi-Dirac) distribution 
$f_n^0=\frac{1}{\exp(p^\mu u_\mu(x)/T)\pm 1}$, $u_\mu(x)$
 is the local fluid four-velocity, and $\tau_{HG}$ is the relaxation 
time (the same for all particle species).
In the local rest frame we have
\begin{equation}
\delta f_n= \frac{\tau_{HG}}{T E}f_n^0\left(1 \pm f_n^0\right) \left(p^i p^j \partial_i v^j-c_s^2 E^2 \partial_i v^j\right) \ .
\end{equation}
Calculating the stress corrections to the  energy momentum tensor
\begin{equation}
\delta T^{\mu \nu}=\pi^{\mu \nu} + \Pi \Delta^{\mu \nu} = \sum_n \int
 \frac{d^3p}{(2\pi)^3 E} p^\mu p^\nu \delta f_n \ ,
\end{equation}
where $\Delta^{\mu\nu}=g^{\mu\nu}- u^\mu u^\nu$,
we  get for the stress tensor in the local rest frame
\begin{equation}
\pi^{i j}=\frac{\tau_{HG}}{T}\sum_n\int\frac{d^3p}{(2\pi)^3}\frac{p^i p^j p^k  p^l}{ E^2}f_n^0\left(1\pm f_n^0\right) \sigma^{k l}
\end{equation}
and
 \begin{equation}
\Pi=\frac{\tau_{HG}}{T}\sum_n\int\frac{d^3p}{(2\pi)^3}\frac{m^3}{3 E^2}f_n^0
\left(1\pm f_n^0 \right) \left(\frac{p^2}{3 E}-c_s^2 E \right) 
 \vec{\nabla}\vec{u}
\end{equation}
with
 \begin{equation}
\sigma_{\alpha\beta}=\frac{1}{2}\left( \nabla_\alpha  u_\beta
+\nabla_\beta u_\alpha -\frac{2}{3}\Delta_{\alpha \beta}\partial_\mu u^\mu\right)\ .
\end{equation}
Comparing with the first order expressions for the stress tensor
$\pi_{\mu\nu}=2\eta \sigma_{\mu\nu}$, $\Pi=-\zeta\partial_\mu u^\mu$ we have
\begin{equation}
\eta=\frac{1}{15T}\sum_n\int \frac{d^3 p}{(2\pi)^3}\frac{p^4}{E^2}f_n^0
\left(1\pm f_n^0 \right)
\label{eq:shdef}
\end{equation}
and
\begin{equation}
\zeta=\frac{\tau_{HG}}{3T}\sum_n\int \frac{d^3 p}{(2\pi)^3}\frac{m^2}{E}f_n^0
\left(1\pm f_n^0 \right)\left(c_s^2 E -\frac{p^2}{3 E}\right) \ .
\label{eq:bulkdef}
\end{equation}

In the modelling of heavy ion collisions we are interested in the properties of the hadronic matter in  a temperature range from 
the freeze-out temperature $T_F>130$MeV to the transition temperature 
$T_c=170$MeV. Performing the sums over the resonances used in the calculation of the hadronic matter equation of state, i.e. the resonances listed in the SHARE program \cite{Torrieri:2004zz}, we can relate the viscosity coefficient 
to the relaxation time. Assuming a constant shear viscosity to entropy ratio 
 $\eta/s=0.1$ between $80$ and $170$MeV,  the relaxation time $\tau_{HG}$
 changes weakly in the range $0.8$-$1.2$fm/c. We could assume instead 
a different 
dependence of $\eta/s$ or of $\tau_{HG}$ on the temperature, but these details 
do not matter much. It turns out that  it is the 
 value at freeze-out that is the most 
important. 
Our choice corresponds to  $\tau_{HG}\simeq 1$fm/c at $T=150$MeV, but 
other values of the parameters
 could be tested in more extensive sets of model calculations. The assumed
 small 
 hadronic shear viscosity is not motivated by microscopic model 
estimates, that would suggest a larger value. It is rather motivated by 
existing viscous hydrodynamic calculations 
 \cite{Romatschke:2007mq,Song:2008hj},
 indicating that the average viscosity in the 
hadronic and QGP phases is small. 

The bulk viscosity of the  hadronic matter at high density is another not 
very well controlled  parameter.
It is expected that in the
 deconfined phase the bulk viscosity coefficient 
is negligible. On the other hand, a sharp rise of the bulk viscosity
has been predicted \cite{Kharzeev:2007wb} around $T_c$. If the effect 
of the bulk viscosity at $T_c$ is large the flow could be modified 
\cite{Song:2009rh,Denicol:2009am} or could even  become 
unstable leading to the fragmentation of the fireball
 \cite{Torrieri:2007fb}. On the other hand,
the  rise of  the bulk viscosity near $T_c$ could be accompanied by 
 critical slowing down, which  leads  to an increase of the dynamical
 bulk viscosity relaxation time $\tau_{\Pi}$, 
delaying the onset and effectively diminishing  bulk viscosity effects.
 $2+1$D hydrodynamic simulations indicate that by the time the 
expanding  system reaches $T_c$ substantial amount of transverse 
flow has already set in \cite{Song:2009rh} and the effects of the rising 
bulk viscosity 
at the critical temperature is reduced and 
  the agreement of the calculation with the data is not spoiled. In this paper
 we do not take into account bulk viscosity near the phase transition.

 In the hadron gas phase the 
bulk viscosity could be quite substantial, 
as particle masses get comparable to the temperature.
Bulk viscosity can be estimated in the relaxation time approximation from Eq. 
(\ref{eq:bulkdef}). The resulting $\zeta/s$  corresponding to $\eta_{HG}/s=0.1$ 
is shown in Fig. \ref{fig:etas}. Since the relaxation time formulas use
 physical hadrons, we restrict the  temperature range for the calculation in 
Eqs. \ref{eq:shdef} and \ref{eq:bulkdef}
 to  the hadronic phase, 
taking
$\frac{\eta}{s}(T)=f_{low}(T)\frac{\eta_{HG}}{s} f_{HG}(T)$.
The shear viscosity at larger temperatures $(1-f_{HG}(T))\frac{\eta_{QGP}}{s}$
is not generated through hadronic processes. In this paper we are interested in
 the effects of bulk viscosity in the late stages, and therefore we 
do not take into account possible bulk viscosity of non-hadronic origin. 
 At temperatures around $150$MeV we 
have $\zeta/s \simeq 0.035$. 
For our estimate of viscosities using a relaxation time formula with $\tau_{HG}$
of the order of $1$fm/c we obtain the  bulk viscosity similar as in microscopic
 models \cite{FernandezFraile:2008vu} but the shear  viscosity is
 significantly smaller than in most estimates 
\cite{Prakash:1993bt,Demir:2008tr,Dobado:2009ek,Chen:2006iga,Itakura:2007mx}.
To check this assumption we performed also a calculation with the same
 bulk viscosity but increasing $\eta_{HG}/s$ to $0.24$. This would mean 
that relaxation time formulas do not apply. We find that the
 assumed value of the shear viscosity 
$\eta_{HG}/s=0.24$, which is still smaller than microscopic estimates,
 gives already a too strong suppression of the elliptic flow.

Eq. (\ref{eq:boltz}) defines 
nonequilibrium corrections to the distribution
 function. The corrections from bulk viscosity cannot be taken in the 
form of the Grad's expansion \cite{Monnai:2009ad}. 
From Eq. (\ref{eq:boltz}) we get for
 the corrections from bulk viscosity $\Pi $ \cite{Sasaki:2008fg}
\begin{equation}
\delta f^{bulk}_n= C_{bulk}f_n^0
\left(1\pm f_n^0 \right)\left(c_s^2 E -\frac{p^2}{3 E}\right) \Pi 
\end{equation}
in the local rest frame,
with 
\begin{equation}
\frac{1}{C_{bulk}}= \frac{1}{3T}\sum_n\int \frac{d^3 p}{(2\pi)^3}\frac{m^2}{E}f_n^0
\left(1\pm f_n^0 \right)\left(c_s^2 E -\frac{p^2}{3 E}\right) \ .
\label{eq:dfbu}
\end{equation} 
The deviation from equilibrium due to the stress corrections
 from shear viscosity are taken in the form \cite{Teaney:2003kp,Baier:2006um}
\begin{equation}
\delta f_{shear}= f_n^0
\left(1\pm f_n^0 \right) \frac{1}{2 T^2 (\epsilon+p)}p^\mu p^\nu \pi_{\mu\nu}
\label{eq:dfsh}
\end{equation}
with $\epsilon$ the local energy density and $p$ the pressure. It must be
 noted that 
more general forms of the nonequilibrium corrections are possible 
for multicomponent systems or for species dependent 
relaxation times \cite{Dusling:2009df}.

\section{Viscous hydrodynamic evolution}

\begin{figure}
\includegraphics[width=.37\textwidth]{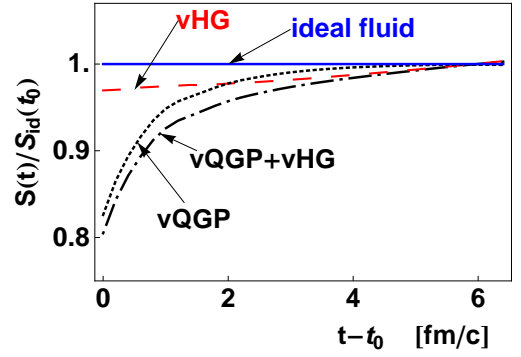}
\caption{(Color online)  Time dependence of the entropy scaled by the 
initial entropy in the ideal fluid calculation. The solid, dashed, 
dash-dotted and dashed 
lines represent results from  the ideal fluid, vQGP, vQGP+vHG
 and vHG calculations respectively.  }
\label{fig:entropy}
\end{figure}

The hydrodynamic equations 
\begin{equation}
\partial_\mu T^{\mu\nu}=0
\end{equation}
are solved in $2+1$dimensions, assuming boost invariance in the 
longitudinal direction. The energy momentum tensor 
\begin{equation}
T^{\mu \nu}= (\epsilon+p)u^\mu u^\nu -p g^{\mu\nu}+\pi^{\mu \nu} + \Pi \Delta^{\mu \nu}
\end{equation}
is composed of the ideal fluid part and stress  shear and bulk viscosity
 corrections
 $\pi$ and $\Pi$. The viscous corrections in the second order Israel-Steward 
viscous hydrodynamics are solutions of the dynamical equations \cite{IS}
\begin{equation}
\Delta^{\mu \alpha} \Delta^{\nu \beta} u^\gamma \partial_\gamma \pi_{\alpha\beta}=\frac{2\eta \sigma^{\mu\nu}-\pi^{\mu\nu}}{\tau_{\pi}}-\frac{1}{2}\pi^{\mu\nu}\frac{\eta T}{\tau_\pi}\partial_\alpha\left(\frac{\tau_\pi u^\alpha}{\eta T}\right) 
\end{equation}
and
\begin{equation}
 u^\gamma \partial_\gamma \Pi=\frac{-\zeta \partial_\gamma u^\gamma-\Pi}{\tau_{\Pi}}-\frac{1}{2}\Pi\frac{\zeta T}{\tau_\Pi}\partial_\alpha\left(\frac{\tau_\Pi u^\alpha}{\zeta T}\right)  \ . \end{equation}
We take for the  relaxation time $\tau_\pi=\frac{3\eta}{T s}$, and assume
$\tau_\Pi=\tau_\pi$. The initial conditions are 
$\pi^{xx}(\tau_0)=\pi^{yy}(\tau_0)=\frac{2\eta}{3\tau_0}$, 
$\pi^{xy}(\tau_0)=0$ and $\Pi(\tau_0)=0$. The details of the choice of initial conditions and $\tau_\Pi$ are not crucial, as 
 the bulk viscosity
 correction gets rapidly  close to the Navier-Stokes value, and anyway 
its influence on the dynamics itself is small.

\begin{figure}
\includegraphics[width=.33\textwidth]{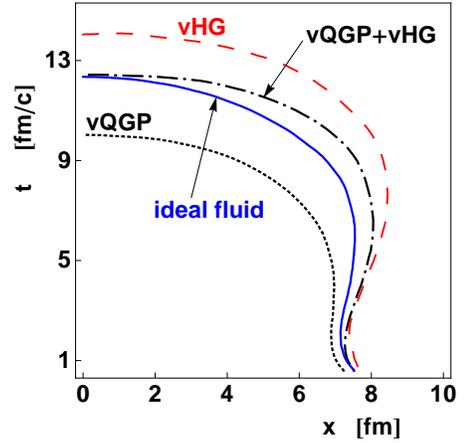}
\caption{(Color online) Freeze-out hypersurfaces $T(t,x,y=0)=T_{F}$ at impact
parameter $b=2.2$fm. The solid, dashed, dash-dotted and dashed 
lines represent hypersurfaces in the ideal fluid, vQGP, vQGP+vHG
 and vHG calculations respectively.}
\label{fig:fr}
\end{figure}

For  the energy density profile in the transverse ($x$-$y$) plane at impact parameter $b$  we use the 
Glauber Model density
\begin{equation}
\epsilon(x,y,b)=\epsilon_0 \frac{ (1-\alpha)\rho_{WN}(x,y,b)+2 \alpha 
\rho_{B}(x,y,b) }{ (1-\alpha)\rho_{WN}(0,0,0)+2 \alpha 
\rho_{B}(0,0,0) }
\label{eq:gm}
\end{equation}
where $\rho_{WN}$ and $\rho_{B}$ are the densities of wounded nucleon and  
 binary collisions respectively, $\alpha=0.145$. The optical Glauber 
Model densities are obtained with Wood-Saxon densities for the Au nuclei
 $\rho_{WS}(r)= \rho_0/\left(\exp\left((r-R_a)/a\right)+1\right)$
 ($\rho_0=0.169$fm$^{-3}$, $R_a=6.38$fm, $a=0.535$fm) and the 
inelastic cross section is $42$mb. The 
energy density at the center of the fireball $\epsilon_0$ for $b=0$ 
 is adjusted to reproduce the particle multiplicity in the most 
central (0-5\%) collisions in ideal hydrodynamic simulations. 
The initial density for other centralities is taken from the formula 
(\ref{eq:gm}) without changing any parameters. 
In viscous hydrodynamic calculations the initial density is rescaled
 to take into account the additional entropy produced.
In Fig. \ref{fig:entropy} is shown the  entropy production in the 
different hydrodynamic evolutions. The entropy is normalized to the 
 entropy in the ideal fluid simulation. In  viscous hydrodynamics the entropy 
increases with time, we chose to normalize the entropy in  all the calculations
 to the same value  at $\tau-\tau_0=6$fm/c. This procedure  yields, 
after hadronization,  similar 
 particle multiplicities in all the calculations. Entropy is produced
 mainly in the QGP phase $\Delta S/S\simeq 20$\%, whereas in the hadronic matter
its relative increase is only  $2$-$3$\%.

The freeze-out temperature is fixed to reproduce the 
transverse momentum spectra of pions in central collisions. %In the cases with
%shear viscosity in the early stages (vQGP and vQGP+vHG) the 
%fluid acquires more transverse flow. Without bulk viscosity at 
%the freeze-out (vQGP) fitting the spectra 
% requires a large freeze-out temperature $150$MeV. 
%Bulk viscosity softens the spectra, and the evolution is continued
% longer to reproduce the  slope of the pion spectrum ($T_F=130$-$135$MeV).
%The ideal fluid evolution case is in between with a freeze-out temperature of 
%$140$MeV.
The lifetime of the fireball is determined by the initial temperature, 
the expansion rate and the freeze-out temperature. The interplay of those 
effects makes the lifetime in the ideal fluid and vQGP+vHG scenarios
 very similar.
 The detailed shape of the freeze-out hypersurface depends however 
on the amount 
of the accumulated transverse flow (Fig. \ref{fig:fr}). 
This has consequences on the resulting
 HBT radii and in particular on the ratio $R_{out}/R_{side}$.

\section{Results}

\begin{figure}
\includegraphics[width=.43\textwidth]{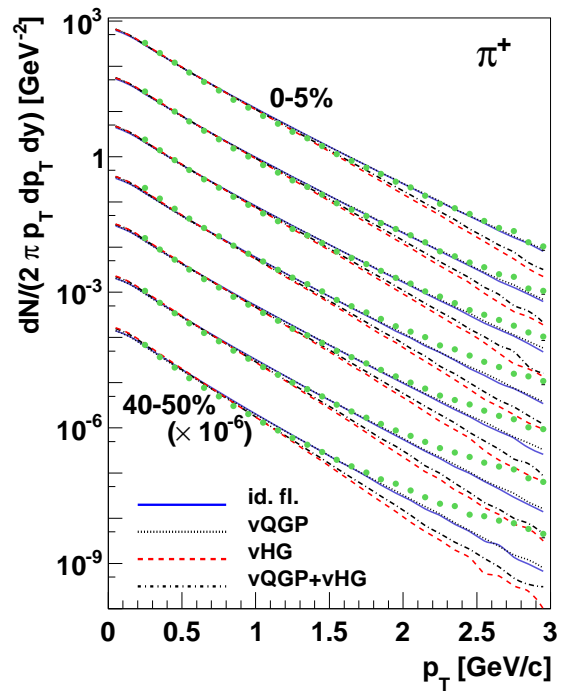}
\caption{(Color online) $\pi^+$ 
transverse momentum spectra spectra for Au-Au collisions at 
$\sqrt{s}=200$GeV and
 centralities $0$-$5$\%, $5$-$10$\%, $10$-$15$\%, $15$-$20$\%,
 $20$-$30$\%, $30$-$40$\% and $40$-$50$\% (successively scaled down 
by powers of $0.1$).
The solid, dotted, dashed and dash-dotted lines represent the 
results of ideal hydrodynamic, vQGP, vHG and vQGP+vHG calculations
 respectively. 
Data are from PHENIX Collaboration \cite{Adler:2003cb}. }
\label{fig:ptpion}
\end{figure}

\begin{figure}
\includegraphics[width=.43\textwidth]{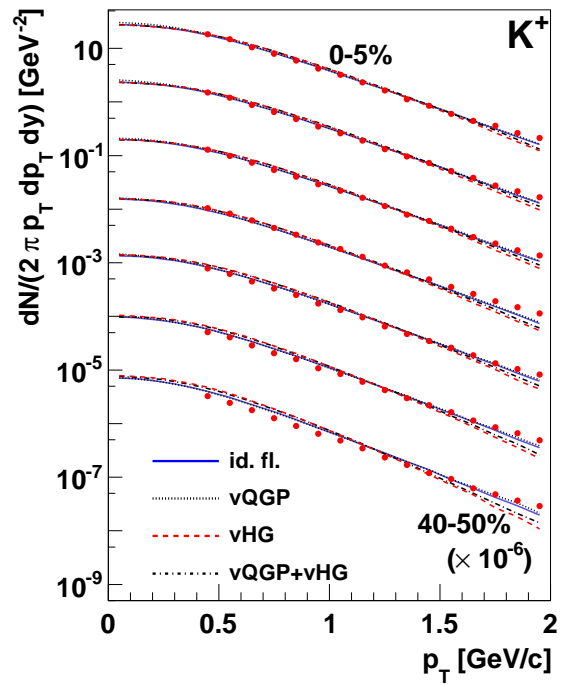}
\caption{(Color online) Same as Fig \ref{fig:ptpion} but for $K^+$. }
\label{fig:ptkaon}
\end{figure}

Transverse momentum spectra of pions are shown in Fig. \ref{fig:ptpion}.
The freeze-out temperature is adjusted for each of the considered scenarios 
to reproduce pion spectra in the most central collisions for $p_\perp<1.2$GeV/c.
In the ideal fluid expansion,
 reducing the freeze-out temperature means that the 
fluid expands longer and more transverse flow builds up. This effect dominates
over the reduction of the final temperature and the spectra become harder. 
For the chosen initial conditions, $T_F=140$MeV is optimal 
for the ideal fluid expanding from $\tau_0=0.6$fm/c.
Shear viscosity corrections  in the plasma phase (scenario vQGP) result in  
additional  transverse pressure in the early stage of the expansion. 
To reproduce the observed pion spectra the evolution must be shortened
 giving  $T_F=150$MeV. The situation is very different if dissipative
 corrections in the hadronic phase are allowed for (scenarios vHG or vQGH+vHG).
Bulk viscosity leads to a substantial softening of light particle 
spectra, hydrodynamic evolution must be effective for a longer time in 
order to reproduce  the $p_\perp$ spectra of pions. Depending on the amount 
of collective transverse flow accumulated in the early phase of the dynamics 
it results in  freeze-out temperatures  $130$-$135$MeV. Bulk viscosity 
corrections (Eq. \ref{eq:dfbu}) grow with the momentum of the particle
 and eventually become as large as the equilibrium distribution $f^0$, it means 
that the formalism breaks down. Using the average  bulk viscosity corrections
 at the freeze-out hypersurface we estimate that viscous hydrodynamics with
 statistical emission of particles breaks down for pion 
momenta of $1.5$GeV/c in the fluid rest frame. Pion spectra at large 
transverse momenta cannot be reliably described in the formalism
 used in this work. After adjusting the freeze-out conditions to reproduce 
pion spectra at soft momenta in central collisions, all observables 
at different 
 centralities are calculated without modifying the parameters of the model.
We observe that pion  spectra at different centralities are well 
described for $p_\perp<1.2$GeV/c.

\begin{figure}
\includegraphics[width=.43\textwidth]{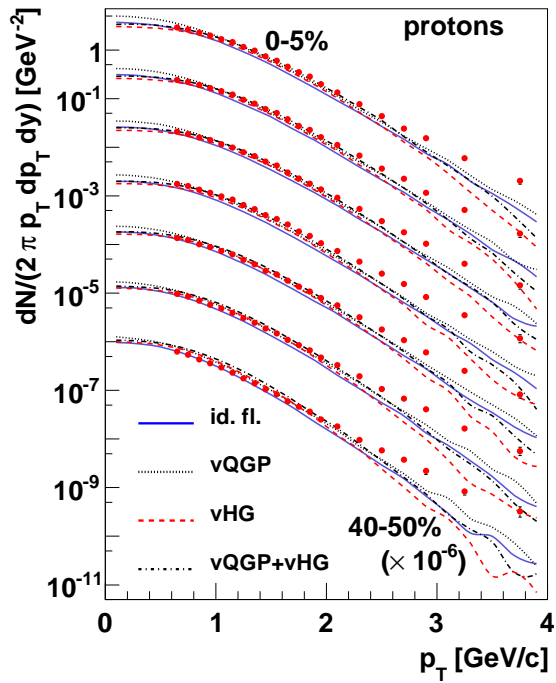}
\caption{(Color online) Same as Fig \ref{fig:ptpion} but for protons. }
\label{fig:ptproton}
\end{figure}

\begin{figure}
\includegraphics[width=.48\textwidth]{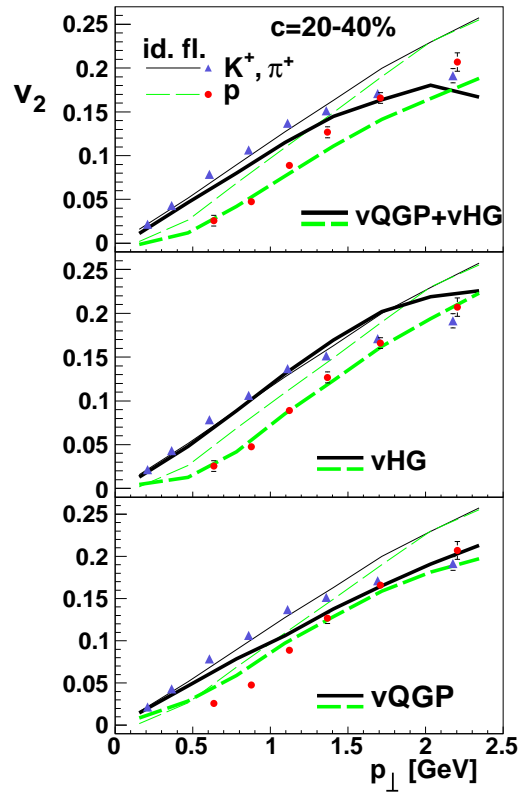}
\caption{(Color online) Elliptic flow coefficient for $\pi^{+}$ and $K^{+}$ 
(triangles and solid lines) and for protons (circles and dashed lines). 
Thin lines are for the ideal fluid model (all panels) and the thick lines for
 the various viscous hydrodynamic calculations, 
data are from the PHENIX Collab. \cite{Adler:2003kt}. 
In the upper, middle and lower panel are shown results obtained 
from vQGP+vHG, vHG and vQGP scenarios  respectively. }
\label{fig:v2pt}
\end{figure}

In Figs. \ref{fig:ptkaon} and \ref{fig:ptproton} we show the spectra
 of $K^+$ and protons at different centralities. A first observation is that
the slopes of the 
spectra for heavier particles obtained in scenarios with
 or without bulk viscosity do not differ as much as for pions. 
It is a consequence of the mass dependence of the bulk viscosity corrections in 
Eq. 
(\ref{eq:dfbu}). Kaon production is 
overpredicted by hydrodynamic calculations in peripheral collisions
which may be a manifestation  of partial equilibration of strangeness 
\cite{Cleymans:2004pp} or of a nontrivial dependence of the thermal 
source size on centrality  \cite{Bozek:2005eu,Becattini:2008ya}.
The effective slopes of proton spectra for $p_\perp<2$GeV/c are 
well reproduced by all the calculations. The multiplicity of protons, 
reflected in the normalization of the spectra in Fig. \ref{fig:ptproton}, 
is better described if bulk viscosity is present. The chemical freeze-out
temperature fitted from the particle number ratios is $165$MeV 
\cite{BraunMunzinger:2001ip,Florkowski:2001fp}, 
significantly
larger than the freeze-out temperatures we use. 
 Simulations where particles are emitted without bulk viscosity
 corrections (id. fl. or vQGP)  underpredict the number of heavier particles. 
Bulk viscosity corrections reduce the number of light particles and lead to
an increase in heavy particle production, resulting in an effective
 chemical non-equilibrium at freeze-out. Consequently, simulations including
 moderate bulk viscosity in the hadronic stage  reproduce the proton number
 in spite of lower freeze-out temperatures.

%\begin{figure}
%\includegraphics[width=.49\textwidth]{v2s.eps}
%\caption{(Color online) Same as in Fig. \ref{fig:v2ch} but scaled but the initial eccentricity from Glauber Model, 
% participant for STAR data and  standard eccentricity for hydrodynamic
% calculations.  }
%\label{fig:v2s}
%\end{figure}

An important characteristic of the dynamics and of the equation of state of the 
fireball is the elliptic flow coefficient \cite{Kolb:2003dz}. 
Most of the elliptic flow is created in the early phase of the expansion,
 and so the flow probes pressure gradients at that time. 
However at  densities where freeze-out occurs the elliptic flow 
is still increasing during the hydrodynamic evolution. Extracting the 
shear viscosity from the comparison of  model calculations to the 
data requires 
a very precise, independent determination of the freeze-out time. 
A strong constraint on the final density at freeze-out is given by the
 transverse momentum dependence of $v_2$ for different species 
\cite{Huovinen:2005gy} and in particular the difference in the flow of 
 pions and protons. This picture is more complicated if viscosity 
corrections at freeze-out are important. First,
 non-equilibrium corrections from shear viscosity 
could 
in principle be very different for different 
particles \cite{Dusling:2009df} and second, bulk viscosity corrections (Eq. 
\ref{eq:dfbu}) depend on the particle mass.

In Fig. \ref{fig:v2pt} we show the momentum dependent elliptic flow 
coefficient for light mesons and protons. The ideal fluid simulation 
(thin lines in all the panels) does not reproduce the meson-proton 
splitting present 
in the data.
The elliptic flow of protons is too large. The same is true for the scenario 
where  the viscosity is negligible in the hadronic phase (lower panel). 
Nonequilibrium corrections at freeze-out (both from shear 
and bulk viscosities) are essentially zero in that case.
Shear viscosity in the plasma phase changes the flow pattern reducing
 velocity gradients and leading to a decrease of the final elliptic flow. 
 Meson and proton elliptic 
flow gets reduced in a similar way by the shear viscosity, and we cannot get
 enough meson-proton splitting. The situation is very different if we allow 
for viscosity  
corrections in the final stage of the expansion. The most 
important difference comes from corrections to the distribution functions at
 freeze-out. Shear viscosity corrections lead to an additional 
reduction of the elliptic flow. However, bulk viscosity corrections reduce 
the transverse momenta of light mesons and the differential elliptic 
flow in $p_\perp$ is increased (two upper panels in Fig. \ref{fig:v2pt}).
The same effect has been noticed in the estimates of bulk viscosity 
corrections at freeze-out in Ref. \cite{Monnai:2009ad}.
Bulk viscosity corrections are much smaller for heavy particles
 and are not sufficient to increase the value of 
 the elliptic flow for protons. 
This and the lower freeze-out temperatures in the  scenarios with hadronic 
bulk viscosity bring   the species dependent elliptic flow to an agreement 
with the data.

\begin{figure}
\includegraphics[width=.45\textwidth]{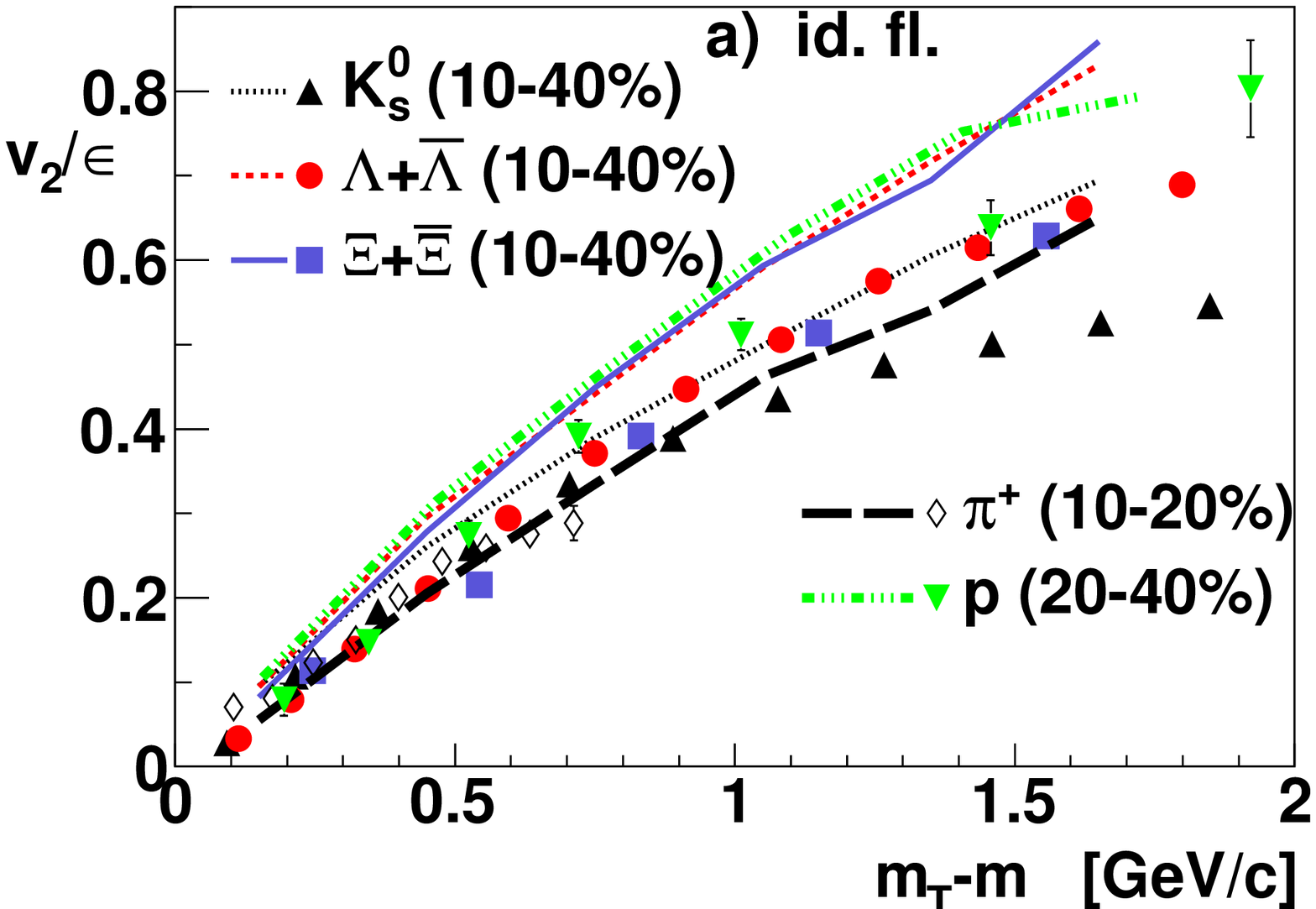}

\includegraphics[width=.45\textwidth]{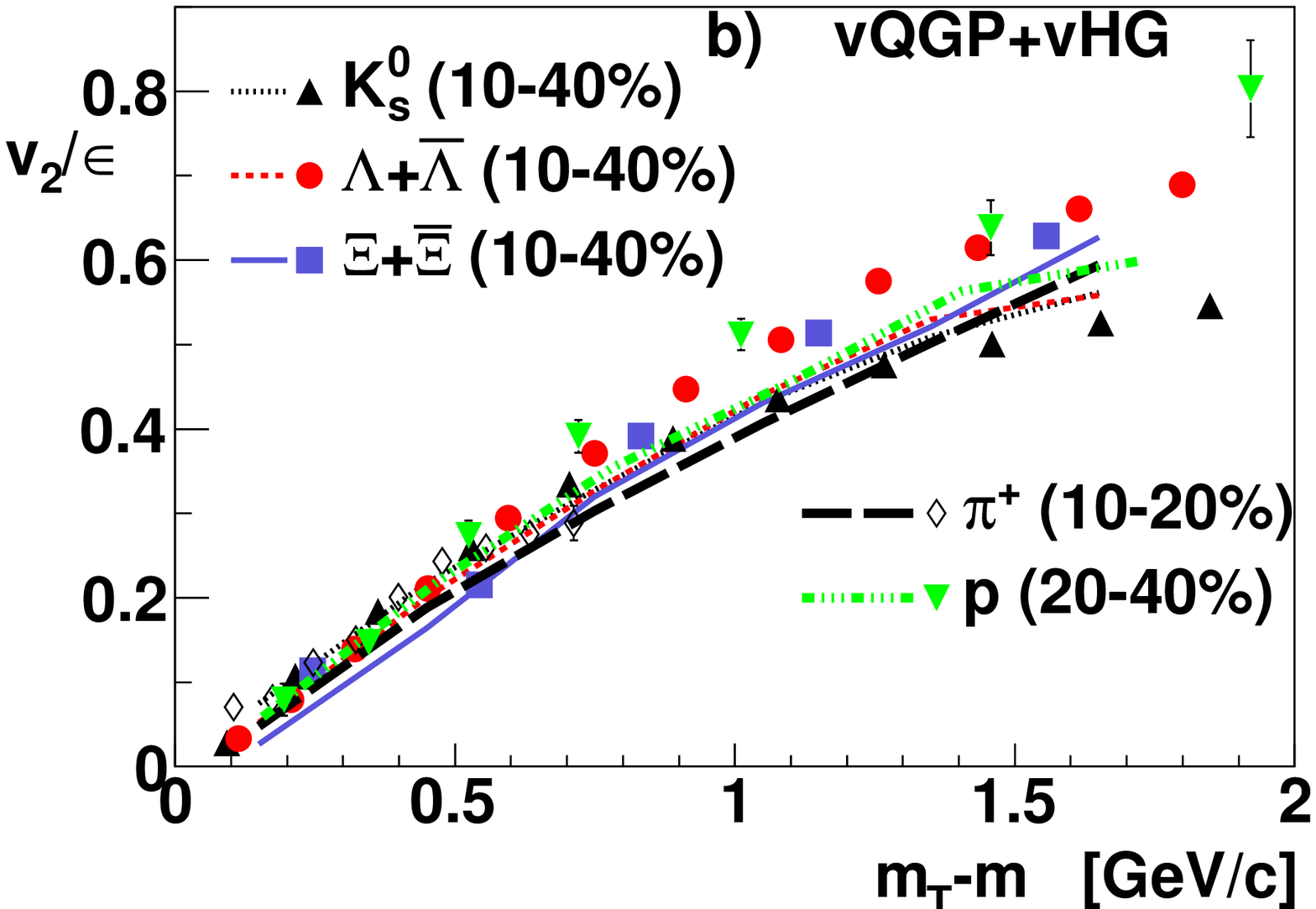}

\includegraphics[width=.45\textwidth]{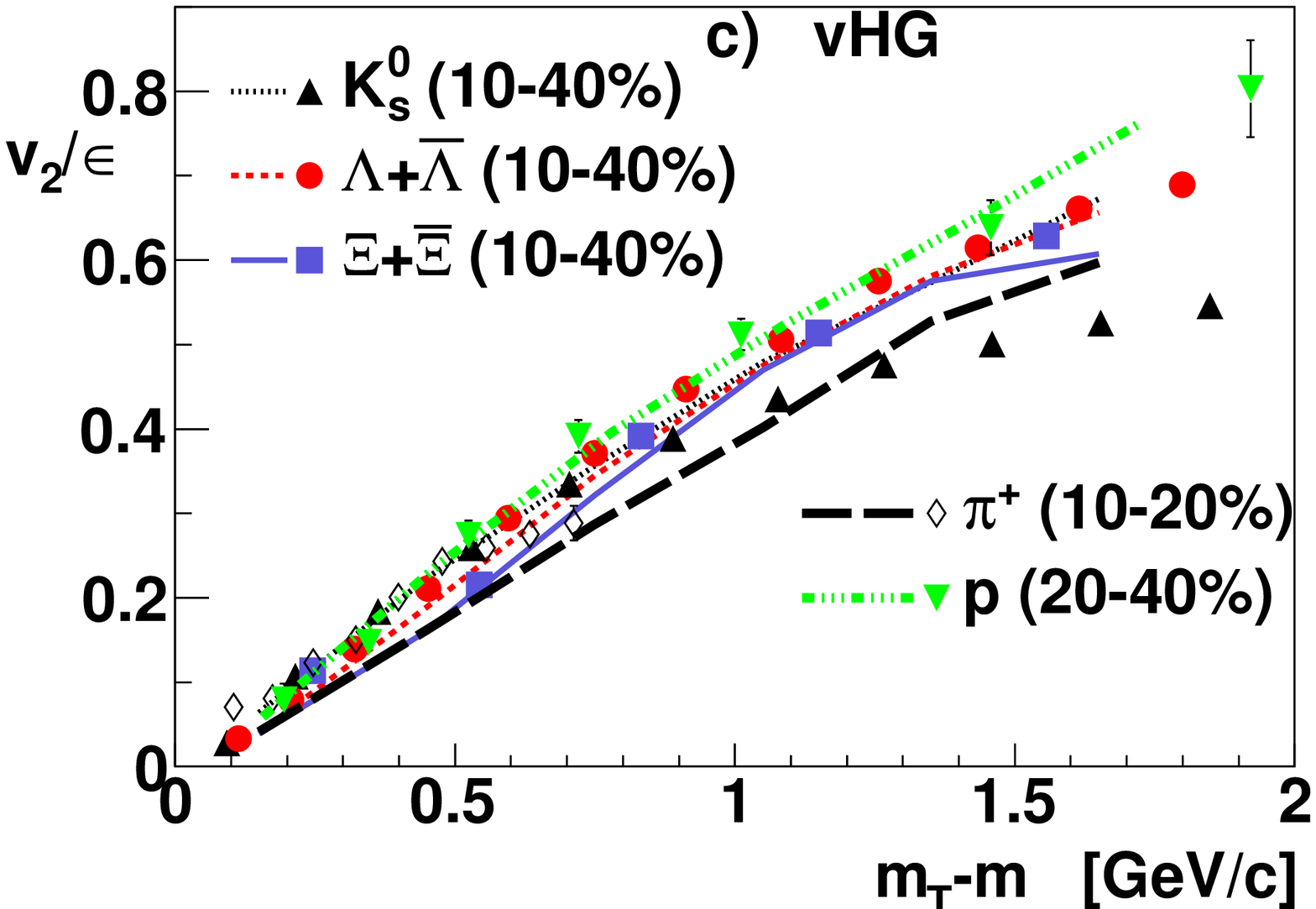}

\includegraphics[width=.45\textwidth]{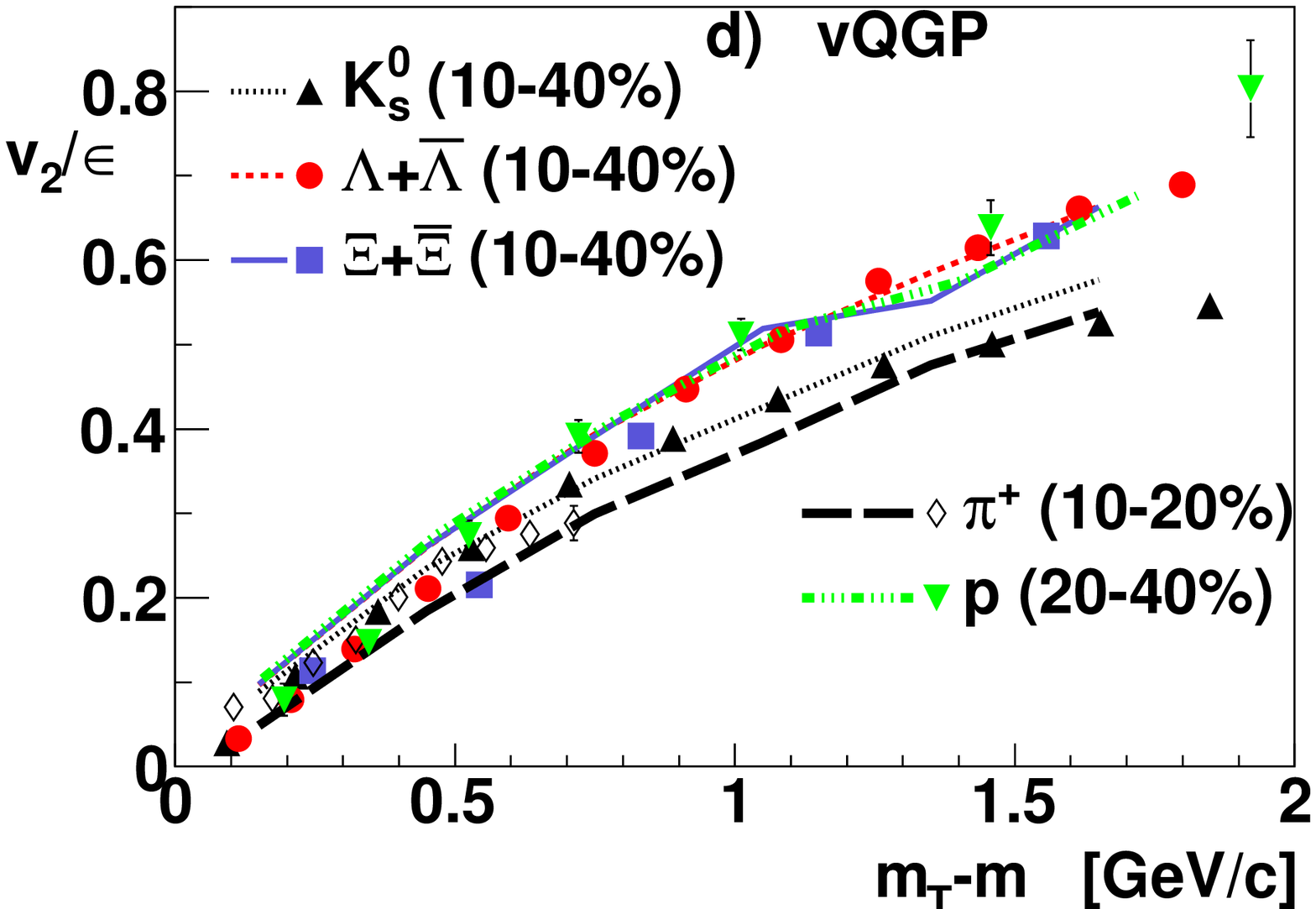}

\caption{(Color online) Elliptic flow as function of 
transverse mass from hydrodynamic calculation 
 (scaled by the initial eccentricity in the calculation, lines) 
and observed experimentally 
(scaled by the participant eccentricity, symbols) for  K$_S^0$ (triangles and 
dotted line), $\Lambda$+$\bar{\Lambda}$ (circles and short dashed line), 
$\Xi$+$\bar{\Xi}$ (squares and solid line), all at centralities $10-40$\%,
 STAR Collab. data \cite{Adams:2005zg}, for $\pi^+$ at centralities $10-20$\%
 (diamonds and long dashed line), PHENIX Collab. data \cite{Adler:2003kt}, 
and for 
protons at centralities $20-40$\% (reversed triangles and dash-dotted line),
 STAR Collab. data \cite{Adams:2004bi}.  }
\label{fig:v2scaling}
\end{figure}

An interesting experimental observation is the mass scaling of identified 
particle elliptic flow in transverse mass \cite{Adams:2005zg} at small
 momenta. The mass ordering of the elliptic flow
 indicates a hydrodynamic origin of the observed flow.  In Fig.
 \ref{fig:v2scaling}  we present the elliptic flow as  function of transverse 
mass for several identified hadrons. The results for different centralities 
are scaled by the initial eccentricity of the fireball.  The results in different
panels correspond to different scenarios for the viscosities. 
The best agreement 
with the transverse mass scaling is seen in the calculations with 
small freeze-out temperatures, vHG or vQGP+vHG. Our calculations in   the 
ideal fluid or vQGP scenarios cannot reproduce the observed mass ordering.
It must be noted that ideal fluid simulations with low freeze-out temperatures
 capture correctly the hydrodynamic origin of the mass ordering of 
the flow \cite{Hirano:2007ei}. Similar results are obtained in hydrodynamics 
with shear viscosity for one value  of $\eta/s$
\cite{Chaudhuri:2009ud}.

\begin{figure}
\includegraphics[width=.47\textwidth]{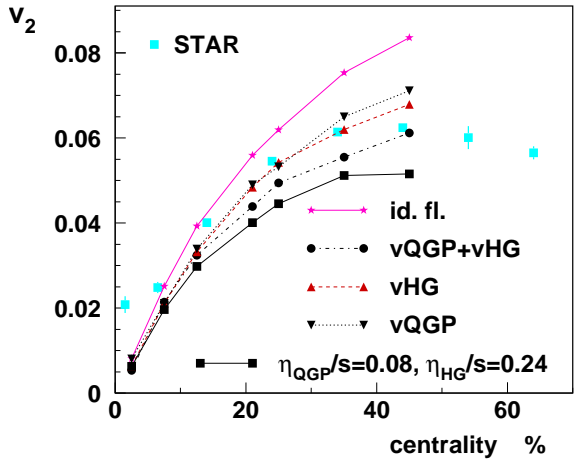}
\caption{(Color online) Elliptic flow coefficient for 
charged particles for different centralities. Stars represent 
ideal fluid results,  reversed triangles, triangles and circles represent the results of  vQGP, vHG and vQGP+vHG  viscous hydrodynamic calculations 
respectively. The solid line with squares denotes the results of  
calculations using $\eta_{HG}/s=0.24$, $\eta_{QGP}/s=0.08$ and the
 same bulk viscosity as in the scenarios vHG and vQGP+vHG.
Data are from the STAR Collab. \cite{Adams:2004bi}. }
\label{fig:v2ch}
\end{figure}

\begin{figure}
\includegraphics[width=.47\textwidth]{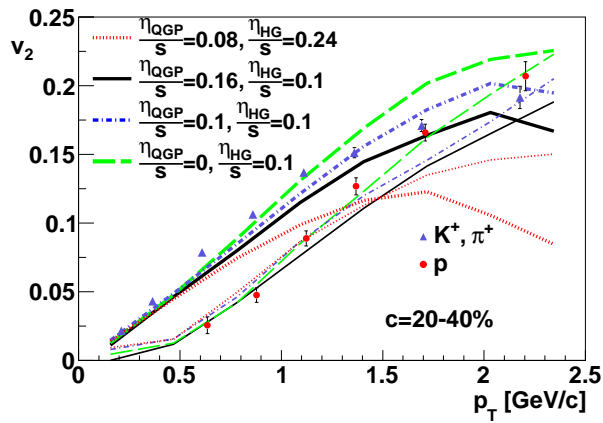}
\caption{(Color online) Elliptic flow of mesons (triangles) and 
protons (circles) as function of transverse momentum \cite{Adler:2003kt},
 compared to 
model calculations. We present three simulations assuming an increasing
  shear viscosity to entropy density ratio with decreasing temperature 
(vHG, $\eta_{QGP}/s=0$, $\eta_{HG}/s=0.1$, dashed lines), a constant one 
($\eta_{QGP}/s=0.1$, $\eta_{HG}/s=0.1$, dashed-dotted lines) and 
a decreasing one (vQGP+vHG, $\eta_{QGP}/s=0.16$, $\eta_{HG}/s=0.1$, solid 
lines). Also are shown results for a calculation with a minimal QGP viscosity 
$\eta_{QGP}/s=0.08$ and moderately large hadronic viscosity $\eta_{HG}/s=0.24$
(dotted lines). Thick and thin  lines represent meson and  proton elliptic flow
respectively.}
\label{fig:v2por}
\end{figure}

In Fig. \ref{fig:v2ch} is plotted the average elliptic flow
 coefficient of charged particles at different centralities.
The 
ideal fluid calculation overpredicts the elliptic flow in peripheral collisions.
The discrepancy
 increases with the impact parameter indicating that corrections to the 
ideal fluid dynamics should be more important in collisions where the hadronic 
phase is relatively more important.  Comparing the three calculations with 
viscosities, we find that adding more dissipative mechanism reduces the final
 elliptic flow. However, the differences between the two scenarios with 
or without viscosity in the plasma phase are not very big.
Most of the effect of the reduction of the azimuthal asymmetry of the flow 
comes from the hadronic dissipation. One must conclude that the sensitivity
 of the elliptic flow to the shear viscosity in the early phase is strongly 
reduced if additional 
dissipation occurs below $T_c$. It must be stressed that the assumed 
strength of shear 
viscosity in the hadronic medium is  small. We performed also simulations 
in a scenario with a larger value of the viscosity in the hadron phase 
$\eta_{HG}/s=0.24$, with $\eta_{QGP}/s=0.08$ and a freeze-out temperature 
of $135$MeV. The result is plotted in Fig. \ref{fig:v2ch} (solid line with
 squares), the elliptic flow of charged particles is below the experimental 
values, 
which means that the assumed shear viscosity is too large. This is in line 
with previous  calculations using a constant value of shear viscosity, where
 the
simulations with  $\eta/s=0.08$ best reproduce the data \cite{Luzum:2008cw}.

In Fig. \ref{fig:v2por} is presented a direct comparison of the elliptic flow 
coefficient as function of transverse momentum for different choices of the 
temperature dependence of the shear viscosity to entropy density ratio.
We take three different values of the QGP viscosity $\eta_{QGP}/s=0,\ 0.1$ and 
$0.16$ and $\eta_{HG}/s=0.1$. It means that we check three qualitatively
 different scenarios with $\eta/s$ increasing, constant or decreasing 
when the temperature drops below $T_c$. The first observation is that  the 
differences between the three calculations are small, moreover 
  increasing $\eta_{QGP}/s$ always leads to a decrease of the elliptic flow.
We find a satisfactory description of the data with a small value of
 $\eta_{HG}/s=0.1$. For  
 a calculation using a minimal QGP shear  
viscosity $\eta_{QGP}/s=0.08$ and a larger value of $\eta_{HG}=0.24$ the 
calculated final elliptic flow of mesons is significantly below the data.
 Increasing $\eta_{QGP}/s$  leads to a decrease 
of $v_2$, also increasing $\eta_{HG}/s$ from $0.1$ to $0.24$ 
gives a strong reduction of the elliptic flow. Comparing these 
results to experimental data  we obtain the following conclusions~:
 both viscosity effects in the plasma and in the hadronic phase 
lead to a decrease of $v_2$, when choosing a small value of
 $\eta_{QGP}/s=0.08$-$0.1$ the best results are obtained for a small 
value of $\eta_{HG}/s=0.1$ (and not $0.24$),  even if the QGP viscosity
 is zero the preferred value of the hadronic viscosity is small.

\begin{figure}
\includegraphics[width=.35\textwidth]{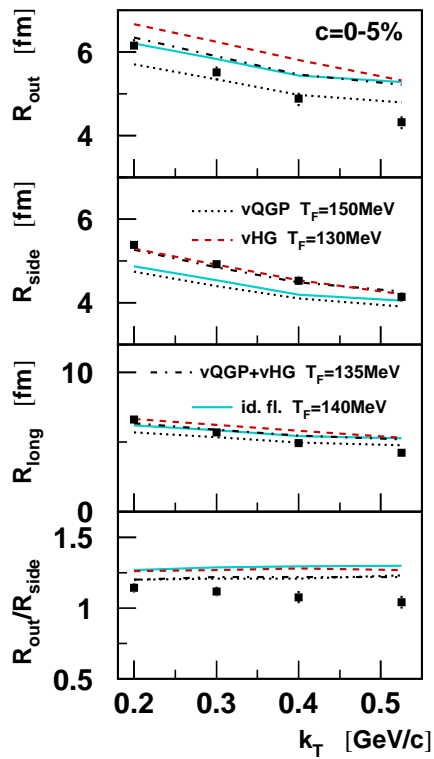}
\caption{(Color online) HBT radii for Au-Au collisions at centrality $0-5\%$.
Ideal fluid  calculation  (solid lines), viscous hydrodynamic
models vQGP+vHG  (dash-dotted lines), vHG (dashed lines), vQGP (dotted lines)
 and STAR Collab. data \cite{Adams:2004yc} (squares) are shown. }
\label{fig:hbt}
\end{figure}

We calculate HBT correlation radii of pions emitted from the fireball 
in the most central collisions. The two particle correlation function 
 sums all pairs of identical pions with interference effects 
\cite{Kisiel:2006is,Kisiel:2006yv}. For a given total momentum
 of the pair $k_\perp$ 
the three-dimensional correlation function in the pion relative
 momentum is fitted with the Bertsch-Pratt formula 
\cite{Pratt:1986ev,Bertsch:1989vn}. All the scenarios of
 the hydrodynamic expansion studied lead to HBT radii that are quite close to 
the data (Fig.\ref{fig:hbt}).  It is a consequence of the hard equation of
 state used, with only a minimal softening around $T_c$ \cite{Broniowski:2008vp}.
The scenarios  differ by the freeze-out temperatures, a
 higher freeze-out temperature means a shorter lifetime and hence smaller 
values of the radii.
$R_{side}$  measuring the geometrical size of the system at freeze-out 
decreases monotonically with increasing
 $T_F$.  The description of the experimentally 
observed small value of the ratio $R_{out}/R_{side}$ requires the use of a 
hard equation of state, early initial time of the expansion, dissipative 
effects and/or 
a Gaussian initial profile \cite{Broniowski:2008vp,Pratt:2008qv}. 
Ideal fluid expansion and the expansion with only hadronic 
dissipation (solid and dashed lines in Fig. \ref{fig:hbt}) give similar results
for $R_{out}/R_{side}$. The 
same is true for scenarios with the same viscosity in 
the early stage but different hadronic dissipation 
(dotted and dash-dotted lines overlap in the lowest 
panel in Fig. \ref{fig:hbt}). We can conclude that the ratio
 $R_{out}/R_{side}$ is sensitive to the early build up of the 
transverse flow and is
 not sensitive to  viscosity effects at freeze-out.
This is in contrast to the elliptic flow which is sensitive to dissipative 
effects at all the stages of the expansion. We cannot reproduce exactly the 
observed HBT radii, this may indicate that the amount of the  early transverse 
flow in the expansion is too low.

%\section{Role of the bulk viscosity}
%
%\begin{figure}
%\includegraphics[width=.49\textwidth]{ptpionynobulk.eps}
%\caption{(Color online) $\pi^+$ 
%transverse momentum spectra spectra as in Fig.
% \ref{fig:ptpion}.
%The solid lines,  dash-dotted and dashed lines represent respectively the 
%results of calculations from  ideal hydrodynamic, vQGP+vHG and vQGP+vHG  with no bulk viscosity. }
%\label{fig:ptpionnobulk}
%\end{figure}

\section{Conclusions}

We present a study  of viscosity effects at 
different stages of the expansion of the fireball created in relativistic 
heavy ion collisions. We introduce the possibility to have two different 
shear viscosities in the QGP and hadronic phases of the matter.  Assuming 
zero  shear viscosity or $\eta/s=0.16$ in the plasma, we test its 
impact on 
 the final observables. The sensitivity of the final elliptic flow observables 
to the early viscosity  is reduced by the dissipative effects in the 
hadronic phase. A crucial effect is the introduction of a moderate value of the 
bulk viscosity in the hadronic medium. Such an assumption is natural in a 
system with partial equilibration and finite particle masses.
The bulk and shear viscosities are treated as free parameters. We 
use however a relaxation time approximation to relate the values of the
 shear and bulk viscosities in the hadron gas. A moderate value of 
the bulk viscosity 
$\zeta/s=0.03$-$0.04$ corresponds to a relatively small value of the shear
 viscosity $\eta/s=0.1$. Bulk viscosity in the late stage leads to a
 shift of the freeze-out temperature to a  value allowing for 
a satisfactory description of the 
mass ordering of the  elliptic flow of identified hadrons without spoiling
 the agreement in the HBT radii and transverse momentum spectra. 

From the 
simulations  using Glauber Model initial densities here presented 
we can conclude that the shear viscosity in the hadronic phase is in the range
$0.1<\eta_{HG}/s<0.24$.  
Shear 
viscosity in the plasma phase  leads to a decrease of elliptic flow, 
but it is the value of $\eta$ at late 
stages that is the most important for  the suppression of the  elliptic flow 
in all the cases $\eta_{QGP}/s<\eta_{HG}/s$, $\eta_{QGP}/s=\eta_{HG}/s$, or
 $\eta_{QGP}/s>\eta_{HG}/s$. Even
 taking $\eta_{QGP}=0$ leads to a preferred value of $\eta_{HG}/s=0.1$
 that best reproduces the data.
Therefore, our results pointing to a small value of the 
shear viscosity in the hadronic phase, are consistent with 
previous calculations using the same 
small value of $\eta$ in the plasma and in the hadron fluids.

The extracted 
 shear viscosity  is significantly below the 
microscopic estimates of the shear viscosity 
in the hadronic matter $\eta_{HG}/s\simeq 1$. 
Using $\eta/s\simeq 1$ in hydrodynamic simulations with the large velocity 
gradients 
in heavy-ion collisions 
is beyond the range 
of applicability of Israel-Steward formalism, it would lead to severe
 numerical instabilities, and is disfavored by elliptic flow data.
It is not clear to the author what is the mechanism that prevents the expected
large  shear viscosity in the hadronic matter  to become 
effective in the viscous hydrodynamic evolution of  heavy ion collisions.
Before concluding that the hadronic matter is indeed a low viscosity fluid, 
it should be checked whether the difference between transport model estimates 
and hydrodynamics is not due to a deficiency of the sudden 
freeze-out procedure used in the calculations
\cite{Huovinen:2008te} or to a strong increase of the relaxation time 
$\tau_\pi$ below $T_c$ analogously to what has been discussed for bulk viscosity
 in Ref. \cite{Song:2009rh}.
%A more extensive study of the parameter set of shear viscosities 
%in the plasma  and hadron phases, and for other models of 
%the initial eccentricity of the fireball will be presented
%in a another work. 

We find that the elliptic flow coefficient is significantly 
reduced due to viscosity effects both in the plasma and in the hadronic matter. 
It means that the extraction of the shear viscosity in QGP is difficult and
 can be reliably addressed only after precisely constraining the 
freeze-out conditions. By this we mean determining both the
 freeze-out temperature and the nonequilibrium effects in the final state.
It is interesting to note that the HBT radii have a simple
 dependence on the choice of the  freeze-out. The radii 
 increase for a larger lifetime of the system, caused by a smaller 
freeze-out temperature. The ratio $R_{out}/R_{side}$ is almost insensitive 
to the freeze-out condition, but it   depends on the amount of the transverse
 flow generated in the early phase. One of the mechanism increasing
 the early transverse flow is due to the shear viscosity in the plasma phase.

Let us close by repeating the observation that  the introduction 
of bulk viscosity in the hadronic medium changes the freeze-out 
conditions  in the  hydrodynamic expansion of the fireball. This allows
for  a good
and simultaneous 
description  of transverse momentum spectra, identified particle 
elliptic flow and HBT radii.

\section*{Acknowledgments}
The author thanks W. Florkowski, U. Heinz, J.-Y. Ollitrault, S. Pratt
and D. Teaney for comments and discussion.
Supported by 
Polish Ministry of Science and Higher Education under
grant N202~034~32/0918
%\appendix*

\bibliography{../hydr}

\begin{thebibliography}{74}
\expandafter\ifx\csname natexlab\endcsname\relax\def\natexlab#1{#1}\fi
\expandafter\ifx\csname bibnamefont\endcsname\relax
  \def\bibnamefont#1{#1}\fi
\expandafter\ifx\csname bibfnamefont\endcsname\relax
  \def\bibfnamefont#1{#1}\fi
\expandafter\ifx\csname citenamefont\endcsname\relax
  \def\citenamefont#1{#1}\fi
\expandafter\ifx\csname url\endcsname\relax
  \def\url#1{\texttt{#1}}\fi
\expandafter\ifx\csname urlprefix\endcsname\relax\def\urlprefix{URL }\fi
\providecommand{\bibinfo}[2]{#2}
\providecommand{\eprint}[2][]{\url{#2}}

\bibitem[{\citenamefont{Arsene et~al.}(2005)}]{Arsene:2004fa}
\bibinfo{author}{\bibfnamefont{I.}~\bibnamefont{Arsene}} \bibnamefont{et~al.}
  (\bibinfo{collaboration}{BRAHMS}), \bibinfo{journal}{Nucl. Phys.}
  \textbf{\bibinfo{volume}{A757}}, \bibinfo{pages}{1} (\bibinfo{year}{2005}).

\bibitem[{\citenamefont{Back et~al.}(2005)}]{Back:2004je}
\bibinfo{author}{\bibfnamefont{B.~B.} \bibnamefont{Back}} \bibnamefont{et~al.}
  (\bibinfo{collaboration}{PHOBOS}), \bibinfo{journal}{Nucl. Phys.}
  \textbf{\bibinfo{volume}{A757}}, \bibinfo{pages}{28} (\bibinfo{year}{2005}).

\bibitem[{\citenamefont{Adams et~al.}(2005{\natexlab{a}})}]{Adams:2005dq}
\bibinfo{author}{\bibfnamefont{J.}~\bibnamefont{Adams}} \bibnamefont{et~al.}
  (\bibinfo{collaboration}{STAR}), \bibinfo{journal}{Nucl. Phys.}
  \textbf{\bibinfo{volume}{A757}}, \bibinfo{pages}{102}
  (\bibinfo{year}{2005}{\natexlab{a}}).

\bibitem[{\citenamefont{Adcox et~al.}(2005)}]{Adcox:2004mh}
\bibinfo{author}{\bibfnamefont{K.}~\bibnamefont{Adcox}} \bibnamefont{et~al.}
  (\bibinfo{collaboration}{PHENIX}), \bibinfo{journal}{Nucl. Phys.}
  \textbf{\bibinfo{volume}{A757}}, \bibinfo{pages}{184} (\bibinfo{year}{2005}).

\bibitem[{\citenamefont{Teaney et~al.}(2001)\citenamefont{Teaney, Lauret, and
  Shuryak}}]{Teaney:2000cw}
\bibinfo{author}{\bibfnamefont{D.}~\bibnamefont{Teaney}},
  \bibinfo{author}{\bibfnamefont{J.}~\bibnamefont{Lauret}}, \bibnamefont{and}
  \bibinfo{author}{\bibfnamefont{E.~V.} \bibnamefont{Shuryak}},
  \bibinfo{journal}{Phys. Rev. Lett.} \textbf{\bibinfo{volume}{86}},
  \bibinfo{pages}{4783} (\bibinfo{year}{2001}).

\bibitem[{\citenamefont{Kolb and Heinz}(2004)}]{Kolb:2003dz}
\bibinfo{author}{\bibfnamefont{P.~F.} \bibnamefont{Kolb}} \bibnamefont{and}
  \bibinfo{author}{\bibfnamefont{U.~W.} \bibnamefont{Heinz}}, in
  \emph{\bibinfo{booktitle}{Quark Gluon Plasma 3}}, edited by
  \bibinfo{editor}{\bibfnamefont{R.}~\bibnamefont{Hwa}} \bibnamefont{and}
  \bibinfo{editor}{\bibfnamefont{X.~N.} \bibnamefont{Wang}}
  (\bibinfo{publisher}{World Scientific, Singapore}, \bibinfo{year}{2004}),
  \eprint{nucl-th/0305084}.

\bibitem[{\citenamefont{Hirano and Tsuda}(2002)}]{Hirano:2002ds}
\bibinfo{author}{\bibfnamefont{T.}~\bibnamefont{Hirano}} \bibnamefont{and}
  \bibinfo{author}{\bibfnamefont{K.}~\bibnamefont{Tsuda}},
  \bibinfo{journal}{Phys. Rev.} \textbf{\bibinfo{volume}{C66}},
  \bibinfo{pages}{054905} (\bibinfo{year}{2002}).

\bibitem[{\citenamefont{Hama et~al.}(2006)}]{Hama:2005dz}
\bibinfo{author}{\bibfnamefont{Y.}~\bibnamefont{Hama}} \bibnamefont{et~al.},
  \bibinfo{journal}{Nucl. Phys.} \textbf{\bibinfo{volume}{A774}},
  \bibinfo{pages}{169} (\bibinfo{year}{2006}).

\bibitem[{\citenamefont{Huovinen and Ruuskanen}(2006)}]{Huovinen:2006jp}
\bibinfo{author}{\bibfnamefont{P.}~\bibnamefont{Huovinen}} \bibnamefont{and}
  \bibinfo{author}{\bibfnamefont{P.~V.} \bibnamefont{Ruuskanen}},
  \bibinfo{journal}{Ann. Rev. Nucl. Part. Sci.} \textbf{\bibinfo{volume}{56}},
  \bibinfo{pages}{163} (\bibinfo{year}{2006}).

\bibitem[{\citenamefont{Hirano et~al.}(2006)\citenamefont{Hirano, Heinz,
  Kharzeev, Lacey, and Nara}}]{Hirano:2005xf}
\bibinfo{author}{\bibfnamefont{T.}~\bibnamefont{Hirano}},
  \bibinfo{author}{\bibfnamefont{U.~W.} \bibnamefont{Heinz}},
  \bibinfo{author}{\bibfnamefont{D.}~\bibnamefont{Kharzeev}},
  \bibinfo{author}{\bibfnamefont{R.}~\bibnamefont{Lacey}}, \bibnamefont{and}
  \bibinfo{author}{\bibfnamefont{Y.}~\bibnamefont{Nara}},
  \bibinfo{journal}{Phys. Lett.} \textbf{\bibinfo{volume}{B636}},
  \bibinfo{pages}{299} (\bibinfo{year}{2006}).

\bibitem[{\citenamefont{Broniowski et~al.}(2008)\citenamefont{Broniowski,
  Chojnacki, Florkowski, and Kisiel}}]{Broniowski:2008vp}
\bibinfo{author}{\bibfnamefont{W.}~\bibnamefont{Broniowski}},
  \bibinfo{author}{\bibfnamefont{M.}~\bibnamefont{Chojnacki}},
  \bibinfo{author}{\bibfnamefont{W.}~\bibnamefont{Florkowski}},
  \bibnamefont{and} \bibinfo{author}{\bibfnamefont{A.}~\bibnamefont{Kisiel}},
  \bibinfo{journal}{Phys. Rev. Lett.} \textbf{\bibinfo{volume}{101}},
  \bibinfo{pages}{022301} (\bibinfo{year}{2008}).

\bibitem[{\citenamefont{Schnedermann et~al.}(1993)\citenamefont{Schnedermann,
  Sollfrank, and Heinz}}]{Schnedermann:1993ws}
\bibinfo{author}{\bibfnamefont{E.}~\bibnamefont{Schnedermann}},
  \bibinfo{author}{\bibfnamefont{J.}~\bibnamefont{Sollfrank}},
  \bibnamefont{and} \bibinfo{author}{\bibfnamefont{U.~W.} \bibnamefont{Heinz}},
  \bibinfo{journal}{Phys. Rev.} \textbf{\bibinfo{volume}{C48}},
  \bibinfo{pages}{2462} (\bibinfo{year}{1993}).

\bibitem[{\citenamefont{Kolb et~al.}(2000)\citenamefont{Kolb, Sollfrank, and
  Heinz}}]{Kolb:2000sd}
\bibinfo{author}{\bibfnamefont{P.~F.} \bibnamefont{Kolb}},
  \bibinfo{author}{\bibfnamefont{J.}~\bibnamefont{Sollfrank}},
  \bibnamefont{and} \bibinfo{author}{\bibfnamefont{U.~W.} \bibnamefont{Heinz}},
  \bibinfo{journal}{Phys. Rev.} \textbf{\bibinfo{volume}{C62}},
  \bibinfo{pages}{054909} (\bibinfo{year}{2000}).

\bibitem[{\citenamefont{Muronga}(2002)}]{Muronga:2001zk}
\bibinfo{author}{\bibfnamefont{A.}~\bibnamefont{Muronga}},
  \bibinfo{journal}{Phys. Rev. Lett.} \textbf{\bibinfo{volume}{88}},
  \bibinfo{pages}{062302} (\bibinfo{year}{2002}).

\bibitem[{\citenamefont{Teaney}(2003)}]{Teaney:2003kp}
\bibinfo{author}{\bibfnamefont{D.}~\bibnamefont{Teaney}},
  \bibinfo{journal}{Phys. Rev.} \textbf{\bibinfo{volume}{C68}},
  \bibinfo{pages}{034913} (\bibinfo{year}{2003}).

\bibitem[{\citenamefont{Baier and Romatschke}(2007)}]{Baier:2006gy}
\bibinfo{author}{\bibfnamefont{R.}~\bibnamefont{Baier}} \bibnamefont{and}
  \bibinfo{author}{\bibfnamefont{P.}~\bibnamefont{Romatschke}},
  \bibinfo{journal}{Eur. Phys. J.} \textbf{\bibinfo{volume}{C51}},
  \bibinfo{pages}{677} (\bibinfo{year}{2007}).

\bibitem[{\citenamefont{Romatschke and Romatschke}(2007)}]{Romatschke:2007mq}
\bibinfo{author}{\bibfnamefont{P.}~\bibnamefont{Romatschke}} \bibnamefont{and}
  \bibinfo{author}{\bibfnamefont{U.}~\bibnamefont{Romatschke}},
  \bibinfo{journal}{Phys. Rev. Lett.} \textbf{\bibinfo{volume}{99}},
  \bibinfo{pages}{172301} (\bibinfo{year}{2007}).

\bibitem[{\citenamefont{Chaudhuri}(2006)}]{Chaudhuri:2006jd}
\bibinfo{author}{\bibfnamefont{A.~K.} \bibnamefont{Chaudhuri}},
  \bibinfo{journal}{Phys. Rev.} \textbf{\bibinfo{volume}{C74}},
  \bibinfo{pages}{044904} (\bibinfo{year}{2006}).

\bibitem[{\citenamefont{Song and Heinz}(2008)}]{Song:2007fn}
\bibinfo{author}{\bibfnamefont{H.}~\bibnamefont{Song}} \bibnamefont{and}
  \bibinfo{author}{\bibfnamefont{U.~W.} \bibnamefont{Heinz}},
  \bibinfo{journal}{Phys. Lett.} \textbf{\bibinfo{volume}{B658}},
  \bibinfo{pages}{279} (\bibinfo{year}{2008}).

\bibitem[{\citenamefont{Israel and Stewart}(1979)}]{IS}
\bibinfo{author}{\bibfnamefont{W.}~\bibnamefont{Israel}} \bibnamefont{and}
  \bibinfo{author}{\bibfnamefont{J.}~\bibnamefont{Stewart}},
  \bibinfo{journal}{Annals Phys.} \textbf{\bibinfo{volume}{118}},
  \bibinfo{pages}{341} (\bibinfo{year}{1979}).

\bibitem[{\citenamefont{Kovtun et~al.}(2005)\citenamefont{Kovtun, Son, and
  Starinets}}]{Kovtun:2004de}
\bibinfo{author}{\bibfnamefont{P.~K.} \bibnamefont{Kovtun}},
  \bibinfo{author}{\bibfnamefont{D.~T.} \bibnamefont{Son}}, \bibnamefont{and}
  \bibinfo{author}{\bibfnamefont{A.~O.} \bibnamefont{Starinets}},
  \bibinfo{journal}{Phys. Rev. Lett.} \textbf{\bibinfo{volume}{94}},
  \bibinfo{pages}{111601} (\bibinfo{year}{2005}).

\bibitem[{\citenamefont{Csernai et~al.}(2006)\citenamefont{Csernai, Kapusta,
  and McLerran}}]{Csernai:2006zz}
\bibinfo{author}{\bibfnamefont{L.~P.} \bibnamefont{Csernai}},
  \bibinfo{author}{\bibfnamefont{J.~I.} \bibnamefont{Kapusta}},
  \bibnamefont{and} \bibinfo{author}{\bibfnamefont{L.~D.}
  \bibnamefont{McLerran}}, \bibinfo{journal}{Phys. Rev. Lett.}
  \textbf{\bibinfo{volume}{97}}, \bibinfo{pages}{152303}
  (\bibinfo{year}{2006}).

\bibitem[{\citenamefont{Song and Heinz}(2009)}]{Song:2008hj}
\bibinfo{author}{\bibfnamefont{H.}~\bibnamefont{Song}} \bibnamefont{and}
  \bibinfo{author}{\bibfnamefont{U.~W.} \bibnamefont{Heinz}},
  \bibinfo{journal}{J. Phys.} \textbf{\bibinfo{volume}{G36}},
  \bibinfo{pages}{064033} (\bibinfo{year}{2009}).

\bibitem[{\citenamefont{Drescher and Nara}(2007)}]{Drescher:2006ca}
\bibinfo{author}{\bibfnamefont{H.~J.} \bibnamefont{Drescher}} \bibnamefont{and}
  \bibinfo{author}{\bibfnamefont{Y.}~\bibnamefont{Nara}},
  \bibinfo{journal}{Phys. Rev.} \textbf{\bibinfo{volume}{C75}},
  \bibinfo{pages}{034905} (\bibinfo{year}{2007}).

\bibitem[{\citenamefont{Kolb et~al.}(2001)\citenamefont{Kolb, Heinz, Huovinen,
  Eskola, and Tuominen}}]{Kolb:2001qz}
\bibinfo{author}{\bibfnamefont{P.~F.} \bibnamefont{Kolb}},
  \bibinfo{author}{\bibfnamefont{U.~W.} \bibnamefont{Heinz}},
  \bibinfo{author}{\bibfnamefont{P.}~\bibnamefont{Huovinen}},
  \bibinfo{author}{\bibfnamefont{K.~J.} \bibnamefont{Eskola}},
  \bibnamefont{and} \bibinfo{author}{\bibfnamefont{K.}~\bibnamefont{Tuominen}},
  \bibinfo{journal}{Nucl. Phys.} \textbf{\bibinfo{volume}{A696}},
  \bibinfo{pages}{197} (\bibinfo{year}{2001}).

\bibitem[{\citenamefont{Miller and Snellings}(2003)}]{Miller:2003kd}
\bibinfo{author}{\bibfnamefont{M.}~\bibnamefont{Miller}} \bibnamefont{and}
  \bibinfo{author}{\bibfnamefont{R.}~\bibnamefont{Snellings}}
  (\bibinfo{year}{2003}), \eprint{nucl-ex/0312008}.

\bibitem[{\citenamefont{Alver et~al.}(2007)}]{Alver:2007rm}
\bibinfo{author}{\bibfnamefont{B.}~\bibnamefont{Alver}} \bibnamefont{et~al.}
  (\bibinfo{collaboration}{PHOBOS}) (\bibinfo{year}{2007}),
  \eprint{nucl-ex/0701049}.

\bibitem[{\citenamefont{Bo\.zek}(2009)}]{Bozek:2008zw}
\bibinfo{author}{\bibfnamefont{P.}~\bibnamefont{Bo\.zek}},
  \bibinfo{journal}{Phys. Rev.} \textbf{\bibinfo{volume}{C79}},
  \bibinfo{pages}{054901} (\bibinfo{year}{2009}).

\bibitem[{\citenamefont{Hirano and Gyulassy}(2006)}]{Hirano:2005wx}
\bibinfo{author}{\bibfnamefont{T.}~\bibnamefont{Hirano}} \bibnamefont{and}
  \bibinfo{author}{\bibfnamefont{M.}~\bibnamefont{Gyulassy}},
  \bibinfo{journal}{Nucl. Phys.} \textbf{\bibinfo{volume}{A769}},
  \bibinfo{pages}{71} (\bibinfo{year}{2006}).

\bibitem[{\citenamefont{Hirano}(2001)}]{Hirano:2000eu}
\bibinfo{author}{\bibfnamefont{T.}~\bibnamefont{Hirano}},
  \bibinfo{journal}{Phys. Rev. Lett.} \textbf{\bibinfo{volume}{86}},
  \bibinfo{pages}{2754} (\bibinfo{year}{2001}).

\bibitem[{\citenamefont{Hirano et~al.}(2007)\citenamefont{Hirano, Heinz,
  Kharzeev, Lacey, and Nara}}]{Hirano:2007xd}
\bibinfo{author}{\bibfnamefont{T.}~\bibnamefont{Hirano}},
  \bibinfo{author}{\bibfnamefont{U.~W.} \bibnamefont{Heinz}},
  \bibinfo{author}{\bibfnamefont{D.}~\bibnamefont{Kharzeev}},
  \bibinfo{author}{\bibfnamefont{R.}~\bibnamefont{Lacey}}, \bibnamefont{and}
  \bibinfo{author}{\bibfnamefont{Y.}~\bibnamefont{Nara}}, \bibinfo{journal}{J.
  Phys.} \textbf{\bibinfo{volume}{G34}}, \bibinfo{pages}{S879}
  (\bibinfo{year}{2007}).

\bibitem[{\citenamefont{Huovinen}(2005)}]{Huovinen:2005gy}
\bibinfo{author}{\bibfnamefont{P.}~\bibnamefont{Huovinen}},
  \bibinfo{journal}{Nucl. Phys.} \textbf{\bibinfo{volume}{A761}},
  \bibinfo{pages}{296} (\bibinfo{year}{2005}).

\bibitem[{\citenamefont{Bass and Dumitru}(2000)}]{Bass:2000ib}
\bibinfo{author}{\bibfnamefont{S.~A.} \bibnamefont{Bass}} \bibnamefont{and}
  \bibinfo{author}{\bibfnamefont{A.}~\bibnamefont{Dumitru}},
  \bibinfo{journal}{Phys. Rev.} \textbf{\bibinfo{volume}{C61}},
  \bibinfo{pages}{064909} (\bibinfo{year}{2000}).

\bibitem[{\citenamefont{Nonaka and Bass}(2007)}]{Nonaka:2006yn}
\bibinfo{author}{\bibfnamefont{C.}~\bibnamefont{Nonaka}} \bibnamefont{and}
  \bibinfo{author}{\bibfnamefont{S.~A.} \bibnamefont{Bass}},
  \bibinfo{journal}{Phys. Rev.} \textbf{\bibinfo{volume}{C75}},
  \bibinfo{pages}{014902} (\bibinfo{year}{2007}).

\bibitem[{\citenamefont{Werner et~al.}(2009)}]{Werner:2009fa}
\bibinfo{author}{\bibfnamefont{K.}~\bibnamefont{Werner}} \bibnamefont{et~al.},
  \bibinfo{journal}{J. Phys.} \textbf{\bibinfo{volume}{G36}},
  \bibinfo{pages}{064030} (\bibinfo{year}{2009}).

\bibitem[{\citenamefont{Prakash et~al.}(1993)\citenamefont{Prakash, Prakash,
  Venugopalan, and Welke}}]{Prakash:1993bt}
\bibinfo{author}{\bibfnamefont{M.}~\bibnamefont{Prakash}},
  \bibinfo{author}{\bibfnamefont{M.}~\bibnamefont{Prakash}},
  \bibinfo{author}{\bibfnamefont{R.}~\bibnamefont{Venugopalan}},
  \bibnamefont{and} \bibinfo{author}{\bibfnamefont{G.}~\bibnamefont{Welke}},
  \bibinfo{journal}{Phys. Rept.} \textbf{\bibinfo{volume}{227}},
  \bibinfo{pages}{321} (\bibinfo{year}{1993}).

\bibitem[{\citenamefont{Demir and Bass}(2009)}]{Demir:2008tr}
\bibinfo{author}{\bibfnamefont{N.}~\bibnamefont{Demir}} \bibnamefont{and}
  \bibinfo{author}{\bibfnamefont{S.~A.} \bibnamefont{Bass}},
  \bibinfo{journal}{Phys. Rev. Lett.} \textbf{\bibinfo{volume}{102}},
  \bibinfo{pages}{172302} (\bibinfo{year}{2009}).

\bibitem[{\citenamefont{Dobado et~al.}(2009)\citenamefont{Dobado,
  Llanes-Estrada, and Torres-Rincon}}]{Dobado:2009ek}
\bibinfo{author}{\bibfnamefont{A.}~\bibnamefont{Dobado}},
  \bibinfo{author}{\bibfnamefont{F.~J.} \bibnamefont{Llanes-Estrada}},
  \bibnamefont{and} \bibinfo{author}{\bibfnamefont{J.~M.}
  \bibnamefont{Torres-Rincon}}, \bibinfo{journal}{Phys. Rev.}
  \textbf{\bibinfo{volume}{D80}}, \bibinfo{pages}{114015}
  (\bibinfo{year}{2009}).

\bibitem[{\citenamefont{Chen and Nakano}(2007)}]{Chen:2006iga}
\bibinfo{author}{\bibfnamefont{J.-W.} \bibnamefont{Chen}} \bibnamefont{and}
  \bibinfo{author}{\bibfnamefont{E.}~\bibnamefont{Nakano}},
  \bibinfo{journal}{Phys. Lett.} \textbf{\bibinfo{volume}{B647}},
  \bibinfo{pages}{371} (\bibinfo{year}{2007}).

\bibitem[{\citenamefont{Itakura et~al.}(2008)\citenamefont{Itakura, Morimatsu,
  and Otomo}}]{Itakura:2007mx}
\bibinfo{author}{\bibfnamefont{K.}~\bibnamefont{Itakura}},
  \bibinfo{author}{\bibfnamefont{O.}~\bibnamefont{Morimatsu}},
  \bibnamefont{and} \bibinfo{author}{\bibfnamefont{H.}~\bibnamefont{Otomo}},
  \bibinfo{journal}{Phys. Rev.} \textbf{\bibinfo{volume}{D77}},
  \bibinfo{pages}{014014} (\bibinfo{year}{2008}).

\bibitem[{\citenamefont{Noronha-Hostler
  et~al.}(2009)\citenamefont{Noronha-Hostler, Noronha, and
  Greiner}}]{NoronhaHostler:2008ju}
\bibinfo{author}{\bibfnamefont{J.}~\bibnamefont{Noronha-Hostler}},
  \bibinfo{author}{\bibfnamefont{J.}~\bibnamefont{Noronha}}, \bibnamefont{and}
  \bibinfo{author}{\bibfnamefont{C.}~\bibnamefont{Greiner}},
  \bibinfo{journal}{Phys. Rev. Lett.} \textbf{\bibinfo{volume}{103}},
  \bibinfo{pages}{172302} (\bibinfo{year}{2009}).

\bibitem[{\citenamefont{Fern\'andez-Fraile and
  Nicola}(2009)}]{FernandezFraile:2008vu}
\bibinfo{author}{\bibfnamefont{D.}~\bibnamefont{Fern\'andez-Fraile}}
  \bibnamefont{and} \bibinfo{author}{\bibfnamefont{A.}
  \bibnamefont{Gomez Nicola}}, \bibinfo{journal}{Phys. Rev. Lett.}
  \textbf{\bibinfo{volume}{102}}, \bibinfo{pages}{121601}
  (\bibinfo{year}{2009}).

\bibitem[{\citenamefont{Chojnacki and Florkowski}(2007)}]{Chojnacki:2007jc}
\bibinfo{author}{\bibfnamefont{M.}~\bibnamefont{Chojnacki}} \bibnamefont{and}
  \bibinfo{author}{\bibfnamefont{W.}~\bibnamefont{Florkowski}},
  \bibinfo{journal}{Acta Phys. Polon.} \textbf{\bibinfo{volume}{B38}},
  \bibinfo{pages}{3249} (\bibinfo{year}{2007}).

\bibitem[{\citenamefont{Chojnacki et~al.}(2008)\citenamefont{Chojnacki,
  Florkowski, Broniowski, and Kisiel}}]{Chojnacki:2007rq}
\bibinfo{author}{\bibfnamefont{M.}~\bibnamefont{Chojnacki}},
  \bibinfo{author}{\bibfnamefont{W.}~\bibnamefont{Florkowski}},
  \bibinfo{author}{\bibfnamefont{W.}~\bibnamefont{Broniowski}},
  \bibnamefont{and} \bibinfo{author}{\bibfnamefont{A.}~\bibnamefont{Kisiel}},
  \bibinfo{journal}{Phys. Rev.} \textbf{\bibinfo{volume}{C78}},
  \bibinfo{pages}{014905} (\bibinfo{year}{2008}).

\bibitem[{\citenamefont{Hosoya and Kajantie}(1985)}]{Hosoya:1983xm}
\bibinfo{author}{\bibfnamefont{A.}~\bibnamefont{Hosoya}} \bibnamefont{and}
  \bibinfo{author}{\bibfnamefont{K.}~\bibnamefont{Kajantie}},
  \bibinfo{journal}{Nucl. Phys.} \textbf{\bibinfo{volume}{B250}},
  \bibinfo{pages}{666} (\bibinfo{year}{1985}).

\bibitem[{\citenamefont{Gavin}(1985)}]{Gavin:1985ph}
\bibinfo{author}{\bibfnamefont{S.}~\bibnamefont{Gavin}},
  \bibinfo{journal}{Nucl. Phys.} \textbf{\bibinfo{volume}{A435}},
  \bibinfo{pages}{826} (\bibinfo{year}{1985}).

\bibitem[{\citenamefont{Sasaki and Redlich}(2009)}]{Sasaki:2008fg}
\bibinfo{author}{\bibfnamefont{C.}~\bibnamefont{Sasaki}} \bibnamefont{and}
  \bibinfo{author}{\bibfnamefont{K.}~\bibnamefont{Redlich}},
  \bibinfo{journal}{Phys. Rev.} \textbf{\bibinfo{volume}{C79}},
  \bibinfo{pages}{055207} (\bibinfo{year}{2009}).

\bibitem[{\citenamefont{Torrieri et~al.}(2005)}]{Torrieri:2004zz}
\bibinfo{author}{\bibfnamefont{G.}~\bibnamefont{Torrieri}}
  \bibnamefont{et~al.}, \bibinfo{journal}{Comput. Phys. Commun.}
  \textbf{\bibinfo{volume}{167}}, \bibinfo{pages}{229} (\bibinfo{year}{2005}).

\bibitem[{\citenamefont{Kharzeev and Tuchin}(2008)}]{Kharzeev:2007wb}
\bibinfo{author}{\bibfnamefont{D.}~\bibnamefont{Kharzeev}} \bibnamefont{and}
  \bibinfo{author}{\bibfnamefont{K.}~\bibnamefont{Tuchin}},
  \bibinfo{journal}{JHEP} \textbf{\bibinfo{volume}{09}}, \bibinfo{pages}{093}
  (\bibinfo{year}{2008}).

\bibitem[{\citenamefont{Song and Heinz}(2010)}]{Song:2009rh}
\bibinfo{author}{\bibfnamefont{H.}~\bibnamefont{Song}} \bibnamefont{and}
  \bibinfo{author}{\bibfnamefont{U.~W.} \bibnamefont{Heinz}},
  \bibinfo{journal}{Phys. Rev.} \textbf{\bibinfo{volume}{{C81}}},
  \bibinfo{pages}{{024905}} (\bibinfo{year}{2010}).

\bibitem[{\citenamefont{Denicol et~al.}(2009)\citenamefont{Denicol, Kodama,
  Koide, and Mota}}]{Denicol:2009am}
\bibinfo{author}{\bibfnamefont{G.~S.} \bibnamefont{Denicol}},
  \bibinfo{author}{\bibfnamefont{T.}~\bibnamefont{Kodama}},
  \bibinfo{author}{\bibfnamefont{T.}~\bibnamefont{Koide}}, \bibnamefont{and}
  \bibinfo{author}{\bibfnamefont{P.}~\bibnamefont{Mota}},
  \bibinfo{journal}{Phys. Rev.} \textbf{\bibinfo{volume}{C80}},
  \bibinfo{pages}{064901} (\bibinfo{year}{2009}).

\bibitem[{\citenamefont{Torrieri et~al.}(2008)\citenamefont{Torrieri, Tomasik,
  and Mishustin}}]{Torrieri:2007fb}
\bibinfo{author}{\bibfnamefont{G.}~\bibnamefont{Torrieri}},
  \bibinfo{author}{\bibfnamefont{B.}~\bibnamefont{Tomasik}}, \bibnamefont{and}
  \bibinfo{author}{\bibfnamefont{I.}~\bibnamefont{Mishustin}},
  \bibinfo{journal}{Phys. Rev.} \textbf{\bibinfo{volume}{C77}},
  \bibinfo{pages}{034903} (\bibinfo{year}{2008}).

\bibitem[{\citenamefont{Monnai and Hirano}(2009)}]{Monnai:2009ad}
\bibinfo{author}{\bibfnamefont{A.}~\bibnamefont{Monnai}} \bibnamefont{and}
  \bibinfo{author}{\bibfnamefont{T.}~\bibnamefont{Hirano}},
  \bibinfo{journal}{Phys. Rev.} \textbf{\bibinfo{volume}{C80}},
  \bibinfo{pages}{054906} (\bibinfo{year}{2009}).

\bibitem[{\citenamefont{Baier et~al.}(2006)\citenamefont{Baier, Romatschke, and
  Wiedemann}}]{Baier:2006um}
\bibinfo{author}{\bibfnamefont{R.}~\bibnamefont{Baier}},
  \bibinfo{author}{\bibfnamefont{P.}~\bibnamefont{Romatschke}},
  \bibnamefont{and} \bibinfo{author}{\bibfnamefont{U.~A.}
  \bibnamefont{Wiedemann}}, \bibinfo{journal}{Phys. Rev.}
  \textbf{\bibinfo{volume}{C73}}, \bibinfo{pages}{064903}
  (\bibinfo{year}{2006}).

\bibitem[{\citenamefont{Dusling et~al.}(2009)\citenamefont{Dusling, Moore, and
  Teaney}}]{Dusling:2009df}
\bibinfo{author}{\bibfnamefont{K.}~\bibnamefont{Dusling}},
  \bibinfo{author}{\bibfnamefont{G.}~\bibnamefont{Moore}}, \bibnamefont{and}
  \bibinfo{author}{\bibfnamefont{D.}~\bibnamefont{Teaney}}
  (\bibinfo{year}{2009}), \eprint{arXiv: 0909.0754 [nucl-th]}.

\bibitem[{\citenamefont{Adler et~al.}(2004)}]{Adler:2003cb}
\bibinfo{author}{\bibfnamefont{S.~S.} \bibnamefont{Adler}} \bibnamefont{et~al.}
  (\bibinfo{collaboration}{PHENIX}), \bibinfo{journal}{Phys. Rev.}
  \textbf{\bibinfo{volume}{C69}}, \bibinfo{pages}{034909}
  (\bibinfo{year}{2004}).

\bibitem[{\citenamefont{Adler et~al.}(2003)}]{Adler:2003kt}
\bibinfo{author}{\bibfnamefont{S.~S.} \bibnamefont{Adler}} \bibnamefont{et~al.}
  (\bibinfo{collaboration}{PHENIX}), \bibinfo{journal}{Phys. Rev. Lett.}
  \textbf{\bibinfo{volume}{91}}, \bibinfo{pages}{182301}
  (\bibinfo{year}{2003}).

\bibitem[{\citenamefont{Cleymans et~al.}(2005)\citenamefont{Cleymans, Kampfer,
  Kaneta, Wheaton, and Xu}}]{Cleymans:2004pp}
\bibinfo{author}{\bibfnamefont{J.}~\bibnamefont{Cleymans}},
  \bibinfo{author}{\bibfnamefont{B.}~\bibnamefont{Kampfer}},
  \bibinfo{author}{\bibfnamefont{M.}~\bibnamefont{Kaneta}},
  \bibinfo{author}{\bibfnamefont{S.}~\bibnamefont{Wheaton}}, \bibnamefont{and}
  \bibinfo{author}{\bibfnamefont{N.}~\bibnamefont{Xu}}, \bibinfo{journal}{Phys.
  Rev.} \textbf{\bibinfo{volume}{C71}}, \bibinfo{pages}{054901}
  (\bibinfo{year}{2005}).

\bibitem[{\citenamefont{Bo\.zek}(2005)}]{Bozek:2005eu}
\bibinfo{author}{\bibfnamefont{P.}~\bibnamefont{Bo\.zek}},
  \bibinfo{journal}{Acta Phys. Polon.} \textbf{\bibinfo{volume}{B36}},
  \bibinfo{pages}{3071} (\bibinfo{year}{2005}).

\bibitem[{\citenamefont{Becattini and Manninen}(2009)}]{Becattini:2008ya}
\bibinfo{author}{\bibfnamefont{F.}~\bibnamefont{Becattini}} \bibnamefont{and}
  \bibinfo{author}{\bibfnamefont{J.}~\bibnamefont{Manninen}},
  \bibinfo{journal}{Phys. Lett.} \textbf{\bibinfo{volume}{B673}},
  \bibinfo{pages}{19} (\bibinfo{year}{2009}).

\bibitem[{\citenamefont{Braun-Munzinger
  et~al.}(2001)\citenamefont{Braun-Munzinger, Magestro, Redlich, and
  Stachel}}]{BraunMunzinger:2001ip}
\bibinfo{author}{\bibfnamefont{P.}~\bibnamefont{Braun-Munzinger}},
  \bibinfo{author}{\bibfnamefont{D.}~\bibnamefont{Magestro}},
  \bibinfo{author}{\bibfnamefont{K.}~\bibnamefont{Redlich}}, \bibnamefont{and}
  \bibinfo{author}{\bibfnamefont{J.}~\bibnamefont{Stachel}},
  \bibinfo{journal}{Phys. Lett.} \textbf{\bibinfo{volume}{B518}},
  \bibinfo{pages}{41} (\bibinfo{year}{2001}).

\bibitem[{\citenamefont{Florkowski et~al.}(2002)\citenamefont{Florkowski,
  Broniowski, and Michalec}}]{Florkowski:2001fp}
\bibinfo{author}{\bibfnamefont{W.}~\bibnamefont{Florkowski}},
  \bibinfo{author}{\bibfnamefont{W.}~\bibnamefont{Broniowski}},
  \bibnamefont{and} \bibinfo{author}{\bibfnamefont{M.}~\bibnamefont{Michalec}},
  \bibinfo{journal}{Acta Phys. Polon.} \textbf{\bibinfo{volume}{B33}},
  \bibinfo{pages}{761} (\bibinfo{year}{2002}).

\bibitem[{\citenamefont{Adams et~al.}(2005{\natexlab{b}})}]{Adams:2005zg}
\bibinfo{author}{\bibfnamefont{J.}~\bibnamefont{Adams}} \bibnamefont{et~al.}
  (\bibinfo{collaboration}{STAR}), \bibinfo{journal}{Phys. Rev. Lett.}
  \textbf{\bibinfo{volume}{95}}, \bibinfo{pages}{122301}
  (\bibinfo{year}{2005}{\natexlab{b}}).

\bibitem[{\citenamefont{Adams et~al.}(2005{\natexlab{c}})}]{Adams:2004bi}
\bibinfo{author}{\bibfnamefont{J.}~\bibnamefont{Adams}} \bibnamefont{et~al.}
  (\bibinfo{collaboration}{STAR}), \bibinfo{journal}{Phys. Rev.}
  \textbf{\bibinfo{volume}{C72}}, \bibinfo{pages}{014904}
  (\bibinfo{year}{2005}{\natexlab{c}}).

\bibitem[{\citenamefont{Hirano et~al.}(2008)\citenamefont{Hirano, Heinz,
  Kharzeev, Lacey, and Nara}}]{Hirano:2007ei}
\bibinfo{author}{\bibfnamefont{T.}~\bibnamefont{Hirano}},
  \bibinfo{author}{\bibfnamefont{U.~W.} \bibnamefont{Heinz}},
  \bibinfo{author}{\bibfnamefont{D.}~\bibnamefont{Kharzeev}},
  \bibinfo{author}{\bibfnamefont{R.}~\bibnamefont{Lacey}}, \bibnamefont{and}
  \bibinfo{author}{\bibfnamefont{Y.}~\bibnamefont{Nara}},
  \bibinfo{journal}{Phys. Rev.} \textbf{\bibinfo{volume}{C77}},
  \bibinfo{pages}{044909} (\bibinfo{year}{2008}).

\bibitem[{\citenamefont{Chaudhuri}(2009)}]{Chaudhuri:2009ud}
\bibinfo{author}{\bibfnamefont{A.~K.} \bibnamefont{Chaudhuri}}
  (\bibinfo{year}{2009}), \eprint{arXiv: 0909.0376 [nucl-th]}.

\bibitem[{\citenamefont{Luzum and Romatschke}(2008)}]{Luzum:2008cw}
\bibinfo{author}{\bibfnamefont{M.}~\bibnamefont{Luzum}} \bibnamefont{and}
  \bibinfo{author}{\bibfnamefont{P.}~\bibnamefont{Romatschke}},
  \bibinfo{journal}{Phys. Rev.} \textbf{\bibinfo{volume}{C78}},
  \bibinfo{pages}{034915} (\bibinfo{year}{2008}).

\bibitem[{\citenamefont{Adams et~al.}(2005{\natexlab{d}})}]{Adams:2004yc}
\bibinfo{author}{\bibfnamefont{J.}~\bibnamefont{Adams}} \bibnamefont{et~al.}
  (\bibinfo{collaboration}{STAR}), \bibinfo{journal}{Phys. Rev.}
  \textbf{\bibinfo{volume}{C71}}, \bibinfo{pages}{044906}
  (\bibinfo{year}{2005}{\natexlab{d}}).

\bibitem[{\citenamefont{Kisiel et~al.}(2006)\citenamefont{Kisiel, Florkowski,
  Broniowski, and Pluta}}]{Kisiel:2006is}
\bibinfo{author}{\bibfnamefont{A.}~\bibnamefont{Kisiel}},
  \bibinfo{author}{\bibfnamefont{W.}~\bibnamefont{Florkowski}},
  \bibinfo{author}{\bibfnamefont{W.}~\bibnamefont{Broniowski}},
  \bibnamefont{and} \bibinfo{author}{\bibfnamefont{J.}~\bibnamefont{Pluta}},
  \bibinfo{journal}{Phys. Rev.} \textbf{\bibinfo{volume}{C73}},
  \bibinfo{pages}{064902} (\bibinfo{year}{2006}).

\bibitem[{\citenamefont{Kisiel}(2007)}]{Kisiel:2006yv}
\bibinfo{author}{\bibfnamefont{A.}~\bibnamefont{Kisiel}},
  \bibinfo{journal}{Braz. J. Phys.} \textbf{\bibinfo{volume}{37}},
  \bibinfo{pages}{917} (\bibinfo{year}{2007}).

\bibitem[{\citenamefont{Pratt}(1986)}]{Pratt:1986ev}
\bibinfo{author}{\bibfnamefont{S.}~\bibnamefont{Pratt}},
  \bibinfo{journal}{Phys. Rev.} \textbf{\bibinfo{volume}{D33}},
  \bibinfo{pages}{72} (\bibinfo{year}{1986}).

\bibitem[{\citenamefont{Bertsch}(1989)}]{Bertsch:1989vn}
\bibinfo{author}{\bibfnamefont{G.~F.} \bibnamefont{Bertsch}},
  \bibinfo{journal}{Nucl. Phys.} \textbf{\bibinfo{volume}{A498}},
  \bibinfo{pages}{173c} (\bibinfo{year}{1989}).

\bibitem[{\citenamefont{Pratt}(2009)}]{Pratt:2008qv}
\bibinfo{author}{\bibfnamefont{S.}~\bibnamefont{Pratt}},
  \bibinfo{journal}{Phys. Rev. Lett.} \textbf{\bibinfo{volume}{102}},
  \bibinfo{pages}{232301} (\bibinfo{year}{2009}).

\bibitem[{\citenamefont{Huovinen and Molnar}(2009)}]{Huovinen:2008te}
\bibinfo{author}{\bibfnamefont{P.}~\bibnamefont{Huovinen}} \bibnamefont{and}
  \bibinfo{author}{\bibfnamefont{D.}~\bibnamefont{Molnar}},
  \bibinfo{journal}{Phys. Rev.} \textbf{\bibinfo{volume}{C79}},
  \bibinfo{pages}{014906} (\bibinfo{year}{2009}).

\end{thebibliography}

\end{document}